\documentclass[twocolumn]{aastex62}

\usepackage[graphicx]{realboxes} 
\usepackage{booktabs}

\accepted{\today}
\submitjournal{ApJ}

\shorttitle{Metallicity study in M83}
\shortauthors{Hernandez et al.}

\begin{document}

\title{\uppercase{First Co-spatial Comparison of Stellar, Neutral-, and Ionized-gas Metallicities in a metal-rich galaxy: M83}\footnote{Based on observations made with the Hubble Space Telescope under program ID 14681.}}

\correspondingauthor{Svea Hernandez}
\email{sveash@stsci.edu}

\author{Svea Hernandez}
\affiliation{AURA for ESA,
Space Telescope Science Institute,
3700 San Martin Drive, 
Baltimore, MD 21218, USA}

\author{Alessandra Aloisi}
\affiliation{Space Telescope Science Institute,
3700 San Martin Drive, 
Baltimore, MD 21218, USA}

\author{Bethan L. James}
\affiliation{AURA for ESA,
Space Telescope Science Institute,
3700 San Martin Drive, 
Baltimore, MD 21218, USA}

\author{Nimisha Kumari}
\affiliation{AURA for ESA,
Space Telescope Science Institute,
3700 San Martin Drive, 
Baltimore, MD 21218, USA}

\author{Danielle Berg}
\affiliation{The Ohio State University,
4061 McPherson Chemical Laboratory,
 140 W 18th Ave., 
 Columbus, OH 43210}
 
\author{Angela Adamo}
\affiliation{Department of Astronomy, 
Oskar Klein Centre, 
Stockholm University, 
AlbaNova University Centre, 
SE-106 91 Stockholm, Sweden}

\author{William P. Blair}
\affiliation{Department of Physics and Astronomy,
The Johns Hopkins University,
3400 N. Charles Street,
Baltimore, MD 21218}

\author{Claude-Andr\'{e}  Faucher-Gigu\`ere}
\affiliation{Department of Physics and Astronomy,
Northwestern University,
2145 Sheridan Road,
Evanston, IL 60208-3112}

\author{Andrew J. Fox}
\affiliation{AURA for ESA,
Space Telescope Science Institute,
3700 San Martin Drive, 
Baltimore, MD 21218, USA}

\author{Alexander B. Gurvich}
\affiliation{Department of Physics \& Astronomy and CIERA, 
Northwestern University, 
1800 Sherman Ave, 
Evanston, IL 60201, USA}

\author{Zachary Hafen}
\affiliation{Department of Physics and Astronomy and Center for Interdisciplinary Exploration and Research in Astrophysics (CIERA), 
Northwestern University, 
2145 Sheridan Road, 
Evanston, IL 60208, USA}

\author{Timothy M. Heckman}
\affiliation{Center for Astrophysical Sciences, 
Department of Physics and Astronomy, 
The Johns Hopkins University, 
Baltimore, MD 21218, USA}

\author{Vianney Lebouteiller}
\affiliation{AIM, CEA, CNRS, 
Universit\'{e} Paris-Saclay, 
Universit\'{e} Paris Diderot, 
Sorbonne Paris Cit\'{e}, 
F-91191 Gif-sur-Yvette, France}

\author{Knox S. Long}
\affiliation{Space Telescope Science Institute,
3700 San Martin Drive, 
Baltimore, MD 21218, USA}
\affiliation{Eureka Scientific, 
Inc. 2452 Delmer Street, Suite 100, 
Oakland, CA 94602-3017, USA}

\author{Evan D. Skillman}
\affiliation{Minnesota Institute for Astrophysics, 
School of Physics and Astronomy,
116 Church Street S.E.,
University of Minnesota, 
Minneapolis, MN 55455, USA}

\author{Jason Tumlinson}
\affiliation{Space Telescope Science Institute,
3700 San Martin Drive, 
Baltimore, MD 21218, USA}
\affiliation{Center for Astrophysical Sciences, 
Department of Physics and Astronomy, 
The Johns Hopkins University, 
Baltimore, MD 21218, USA}

\author{Bradley C. Whitmore}
\affiliation{Space Telescope Science Institute,
3700 San Martin Drive, 
Baltimore, MD 21218, USA}

\begin{abstract}
We carry out a comparative analysis of the metallicities from the stellar, neutral-gas, and ionized-gas components in the metal-rich spiral galaxy M83. We analyze spectroscopic observations taken with the Hubble Space Telescope (HST), the Large Binocular Telescope (LBT) and the Very Large Telescope (VLT). We detect a clear depletion of the \ion{H}{1} gas, as observed from the \ion{H}{1} column densities in the nuclear region of this spiral galaxy. We find column densities of log[$N$(\ion{H}{1}) cm$^{-2}$] $<$ 20.0 at galactocentric distances of $<$ 0.18 kpc, in contrast to column densities of log[$N$(\ion{H}{1}) cm$^{-2}$] $\sim$ 21.0 in the galactic disk, a trend observed in other nearby spiral galaxies. We measure a metallicity gradient of $-$0.03 $\pm$ 0.01 dex kpc$^{-1}$ for the ionized gas, comparable to the metallicity gradient of a local benchmark of 49 nearby star-forming galaxies of  $-$0.026 $\pm$ 0.002 dex kpc$^{-1}$. Our co-spatial metallicity comparison of the multi-phase gas and stellar populations shows excellent agreement outside of the nucleus of the galaxy hinting at a scenario where the mixing of newly synthesized metals from the most massive stars in the star clusters takes longer than their lifetimes ($\sim$10 Myr). Finally, our work shows that caution must be taken when studying the metallicity gradient of the neutral-gas component in star-forming galaxies, since this can be strongly biased, as these environments can be dominated by molecular gas. In these regions the typical metallicity tracers can provide inaccurate abundances as they may trace both the neutral- and molecular-gas components. 
 \end{abstract}

\keywords{galaxies -- abundances -- ISM -- starburst}

\section{Introduction}\label{sec:intro}
Understanding the formation and evolution of galaxies continues to be one of the main quests in modern astrophysics. Extragalactic abundance measurements have greatly contributed to uncover a variety of physical and evolutionary processes influencing events taking place within and among galaxies. Studies of galactic gradients and global metallicity relations (i.e., mass-metallicity relation, MZR) are widely used to investigate star-formation episodes, galactic winds, and accretion of pristine matter \citep{sea71,mcc85,zar94,tre04,and13,kud15}. \par
The measurements from both the MZR and metallicity trends in star-forming galaxies (SFGs), have relied for decades  on the analysis of emission lines from \ion{H}{2} regions. Typically, when inferring the gas-phase metallicities from the most common heavy element, oxygen, two main techniques are applied: strong-line analysis, and the electron temperature based ($T_{e}$-based) method. The strong-line analysis is based on the flux ratios of some of the strongest forbidden lines, e.g.,  [\ion{O}{2}] and [\ion{O}{3}], with respect to hydrogen \citep{pag79, mcg94}. On the other hand, the ``direct" method, or $T_{e}$-based method, relies on the flux ratio of auroral to nebular lines, e.g., [\ion{O}{3}] $\lambda$4363/[\ion{O}{3}] $\lambda\lambda$4959, 5007, to measure the temperature of the high-excitation zone \citep{din85,rub94,lee04, sta05}. Furthermore, in the last decades these nebular techniques have been extended to studies of star-forming galaxies at high redshift \citep{pet01,kob04,sav05,cow08}. Although the analysis of \ion{H}{2} regions have made invaluable contributions when it comes to investigating the present-day chemical state of starburst galaxies, these regions could be enhanced compared to the surrounding interstellar medium \citep[ISM,][]{kun86, leb13}. And only in a few cases (e.g., \citealp{san15, lag18}) studies have found the metallicities in \ion{H}{2} regions to be lower than the rest of the galaxy. This has been attributed to infall of cold metal-poor gas. \par
An alternative method for investigating the chemical composition of galaxies is to directly study the neutral gas in SFGs. The metal content of a galaxy can be examined through the analysis of the absorption lines in their far ultraviolet (UV) spectroscopic observations. A common technique is to use bright UV targets within these galaxies as background sources \citep{kun94}. In such observations, the metals along the line of sight imprint absorption features on the UV continuum of such targets. This technique has been applied extensively to local galaxies \citep{alo03,leb13,wer13,jam14}. This approach not only allows us to study the metal contents accounting for the bulk of the mass of the galaxy, it also provides us with a view of the metal enrichment over large spatial scales and long timescales (dilution of abundances in \ion{H}{1} regions).  \par

A third approach to studying the metallicities of star-forming galaxies is the analysis of young stellar populations or individual stars. New techniques have been developed in the last decade to investigate the chemical contents of nearby galaxies using blue supergiants, BSGs, and red supergiants, RSGs, as metallicity tracers \citep{dav10,dav15,dav17,dav17b,kud12,kud13,kud14,hos14}. Additionally, it is also possible to measure stellar metallicities from integrated-light spectroscopic observations of star clusters in nearby galaxies using high-resolution observations \citep{col11,col12,lar06,lar08,lar12,lar14,lar17,lar18}. This same technique has most recently been extended to intermediate-resolution observations of extragalactic stellar populations \citep{her17,her18a,her18b,her19} as well as for populations at high redshifts \citep{hal08, ste16,chi19}. \par

In spite of the variety of tools available to investigate the chemical composition of star-forming galaxies, detailed comparisons of the abundances obtained from the ionized-gas, neutral-gas, and stellar components are needed to fully understand the chemical state and evolution of galaxies. The general expectation is that young populations of stars should have a similar chemical composition as their parent gas cloud and associated \ion{H}{2} region. In this context, several studies in low-metallicity and chemically homogenous environments have shown agreement between the ionized-gas and young stellar population metallicities \citep{bre06,lee06}. Studies of different galaxies with higher metallicities, such as spiral galaxies, have shown a varying degree of agreement between their nebular and stellar metallicities, from excellent ($<$0.1 dex) to differences as high as $\sim$0.2 dex \citep{bre09, bre16, hos14, dav17b}. Even more intriguing is the fact that different studies have also hinted that for high-metallicity environments the T$_{e}$-based method underestimates the metallicities; this is suggested when compared to stellar abundances \citep{zur12, sim11,gar14}. These high-metallicity environments are particularly challenging to study as the application of the direct method is limited given that the temperature-sensitive lines are typically too weak to be detected \citep{bre05}. \par

Studies comparing different metallicity diagnostics show conflicting results. Through a study of 30 SFGs at z$\sim$2, \citet{ste16} found a factor of $\sim$4-5 difference between their inferred stellar and nebular metallicities, with a clear enhancement in the metallicities of the ionized gas. They argue that these artificially low stellar metallicities are observed due to the supersolar $\alpha$/Fe abundance ratios of these galaxies at z$\sim$2. They assume that this behavior is expected at high redshift, and might be rare in low-$z$ environments due to systematic differences in the star-formation history of typical galaxies. In a more recent study, \citet{chi19} measure the metallicity of these same components, ionized-gas and stellar, in a sample of 61 SFGs at z $< $0.2, and 19 galaxies at z$\sim$ 2. In contrast to the work by \citet{ste16}, \citet{chi19} conclude that the stellar and nebular metallicities are similar to each other when assuming mixed-age stellar populations. Under the general assumption that young stellar populations have a similar chemical composition as their parent gas cloud and associated \ion{H}{2} region, the work by \citet{chi19} hints at a scenario where the gas surrounding high-mass stars is not instantaneously metal-enriched by massive stars. This would imply that increasing the metallicity of the adjacent interstellar gas takes longer than the inferred lifetimes ($\sim$10 Myr) of the massive stellar populations.\par
In this paper we analyze observations from the Cosmic Origin Spectrograph (COS) on Hubble, as well as data from the Multi-Object Double Spectrograph (MODS) on the Large Binocular Telescope (LBT) and the Multi Unit Spectroscopic Explorer (MUSE) on the Very Large Telescope (VLT) for a sample of pointings distributed across the face of the metal-rich spiral galaxy M83 (NGC 5236). M83 is our nearest face-on grand-design spiral and starburst galaxy \citep{dop10} at a distance of 4.9 Mpc \citep[][derived from the magnitudes of the tip of the red giant branch, TRGB]{jac09}. Its proximity and orientation allow for a spatially resolved study of its different components: stellar, neutral gas, and ionized gas. We present in Table \ref{table:gal} a detailed list of the general parameters of M83. Our main motivation is to understand how abundances from the different galaxy components relate to each other, particularly in a challenging environment as is the metal-rich regime. In Section \ref{sec:obs} we describe the observations and data reduction. The analysis of the different observations and for the different metallicity components is detailed in Section \ref{sec:ana}. We discuss our findings in Section \ref{sec:discussion}, and provide our concluding remarks in Section \ref{sec:con}.

\begin{table}
\caption{General parameters for M83}
\label{table:gal}
\centering 
\begin{tabular}{lc}
\hline \hline
Parameter & Value \\
\hline\\
R.A. (J2000.0) & 204.253958$^{\circ}$\\
Dec (J2000.0) & -29.865417$^{\circ}$\\
Distance$^{a}$ & 4.9 Mpc\\
Morphological type & SAB(s)c \\
$R_{25}\:^{b}$ & 6.44$\arcmin\;$  (9.18 kpc)\\
Inclination$^{b}$ & 24\degr\\
Position Angle$^{c}$ & 45\degr\\
Heliocentric radial velocity & 512.95 km s$^{-1}$\\
\hline
\end{tabular}
\begin{minipage}{15cm}~\\
\textbf{Notes.} All parameters from the NASA Extragalactic \\Database (NED), except where noted.\\
 \textsuperscript{$a$}{\citet{jac09}} \\
 \textsuperscript{$b$}{\citet{dev91}} \\
 \textsuperscript{$c$}{\citet{com81}}\\
 \end{minipage}
\end{table}

\section{Observations and Data Reduction}\label{sec:obs}
\subsection{COS observations}
The analysis done here relies on the observations taken as part of HST program ID (PID) 14681 (PI: Aloisi), collected between 2017 May--July. The targets were acquired using near-UV (NUV) imaging, and the spectroscopic data were observed with the G130M/1291 and G160M/1623 settings returning wavelength coverage from 1130 to 1800 \r{A}. The data were collected at Lifetime Position 3 providing a wavelength-dependent resolution ranging between $R\sim$15,000 and 20,000. 
The targets observed in HST program ID 14681 were chosen from the list of young star clusters in the Wide Field Camera 3 (WFC3) Early Release Science Cycle 17 GO/DD PID 11360 (PI: O'Connell) and Cycle 19 GO PID 12513 (PI: Blair). An overview of this WFC3 multi-wavelength campaign is provided in \citet{bla14}. Our selection criterion required targets to have magnitudes $m_{\rm F336W}\lesssim$ 17. We list in Table \ref{table:obs} the properties of our COS sample along with the information of their observations. We note that the spectroscopic observations for the M83-5 pointing were affected by a guide star filure after the science exposures began collecting data. This caused the S/N of the resulting spectroscopic data for this target to be lower than originally planned. For this reason we have excluded this target from our analysis. \par
\begin{figure*}
   	  \includegraphics[scale=0.155]{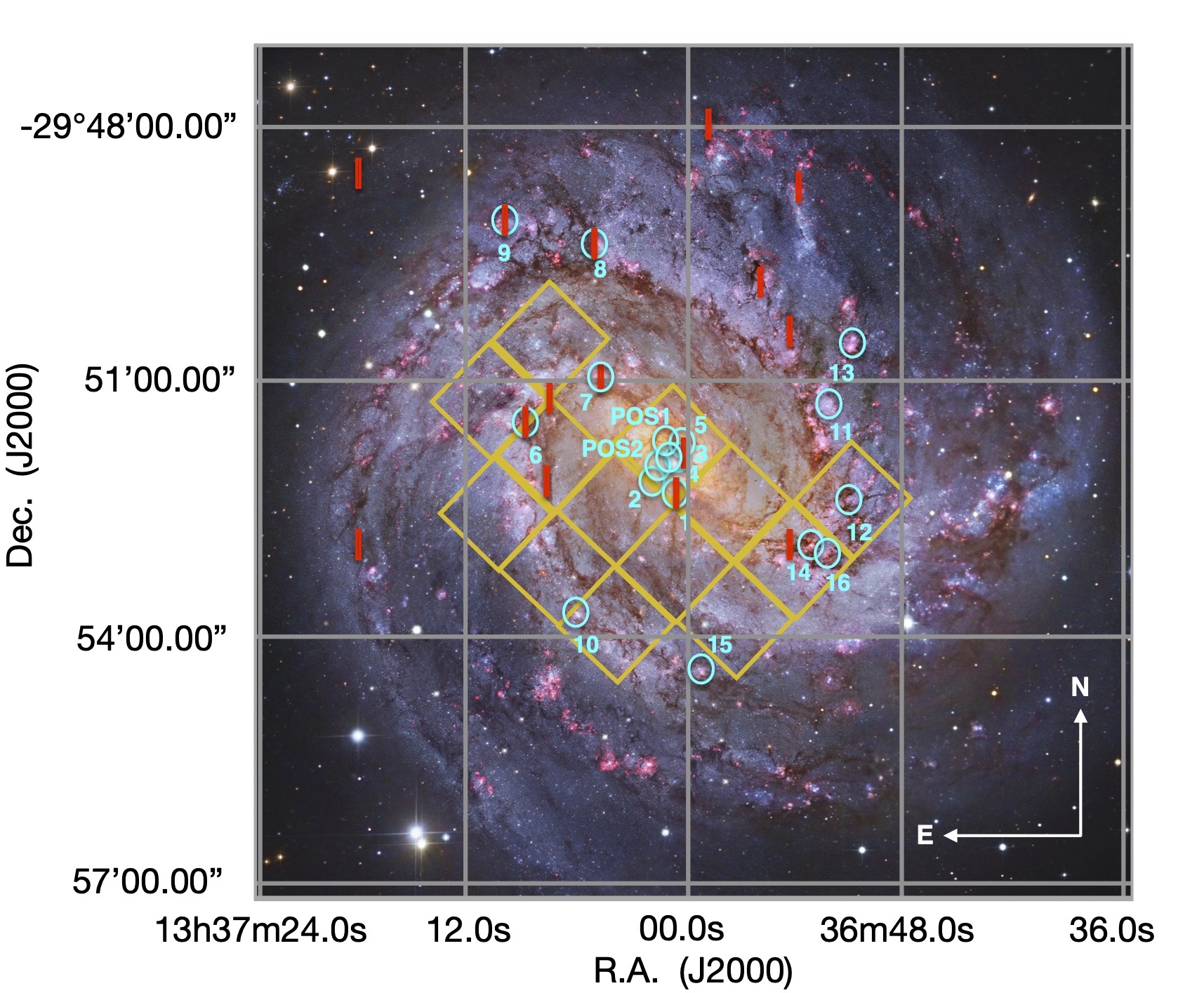}
	  \includegraphics[scale=0.405]{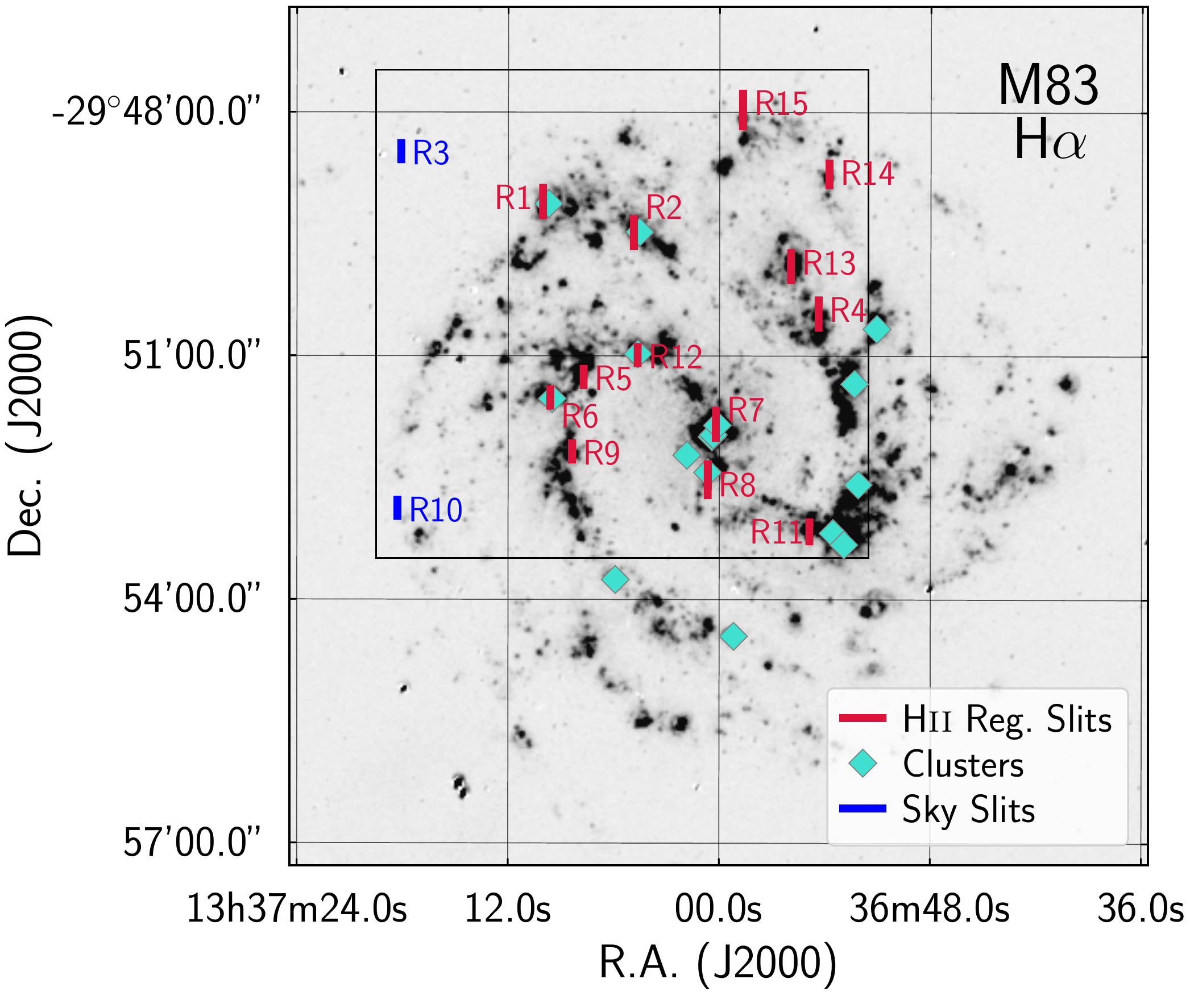}
      \caption{ $Left$: Color-composite image observed with the 2.2m Max Planck-ESO Telescope, the 8.2m Subaru Telescope (NAOJ), and the Hubble Space Telescope. We marked with cyan circles the location of the star clusters observed with COS. We overlay the footprints of the MODS slits in red showing the position of the observed \ion{H}{2} regions. MUSE FoV are shown in yellow. Processing and Copyright: Robert Gendler. $Right$: Spitzer Local Volume Legacy Survey H$\alpha$ image of M83 \citep{dale09}. 
The footprint of the LBT/MODS mask is shown as a gray square, with sky slits (blue) and \ion{H}{2} region slit locations (red) overlaid in comparison to the stellar clusters targeted with COS (cyan diamonds).
The slit positions targeted \ion{H}{2} regions associated with YMCs, although additional associated \ion{H}{2} 
regions were added in order to maximize effective usage of mask real estate.}
         \label{fig:m83}
   \end{figure*}

In addition to the targets from PID 14681, we extended our analysis to include two more COS pointings, M83-POS1 and M83-POS2, observed as part of HST PIDs 11579 and 15193 (PI: Aloisi). These additional observations were taken using the G130M/1291 and G160M/1623 setting with similar wavelength coverage as that from PID 14681. Our final COS sample is composed of 17 pointings distributed throughout the disk of M83 as indicated in Figure \ref{fig:m83}. \par
We retrieved the observations from the Mikulski Archive for Space Telescopes (MAST) and calibrated them using the HST pipeline, \textsc{CALCOS v3.3.4} \citep{fox15}. More details on the reduction of the observations are provided by \citet{her19}. As a final step we bin the spectra by a COS resolution element (1 resel = 6 pixels), corresponding to the nominal point-spread function. \par

\begin{table*}
\caption{Properties of the observed targets and their COS Observations}
\label{table:obs}
\centering 
\begin{tabular}{cccccccccccc}
\hline \hline
Cluster & R.A.$^{\dagger}$ & Dec$^{\dagger}$ & $m_{\rm F336W}^{a}$ & $R/R_{25}$ $^{b}$ & \multicolumn{2}{c}{$t_{\rm exp}$ (s)} & \multicolumn{2}{c}{S/N $^c$ (resel$^{-1}$)}   \\
 & \textbf{($^{\circ}$)} & \textbf{($^{\circ}$)} & (mag) & & G130M & G160M &  G130M & G160M\\
\hline\\
M83-1 & 204.2527583&-29.8739111 & 17.00 &  0.08 &2368 & 8114  & 1.5 & 1.3  \\
M83-2 & 204.2576792&-29.8703833 & 16.14 &  0.06 &2120 & 5652  & 2.2 & 1.5 \\
M83-3 & 204.2517375&-29.8666222 & 15.10 &  0.02 &400 & 1304  & 6.9 & 5.5 \\
M83-4 & 204.2514333&-29.8662056 & 14.40 &  0.02 & 500 & 1208  & 3.1 & 2.2 \\
M83-5 &204.2504500&-29.8642500 & 14.85 &  0.04 & 1600 & 3072  & 0.9  & 0.6\\
M83-6 & 204.2895625&-29.8588083 & 16.46 &  0.34 & 1036 & 3645  & 4.0 & 3.5 \\
M83-7 & 204.2692667&-29.8495917 & 14.79 &  0.21 &540 & 1136 & 2.2 & 1.3\\
M83-8 & 204.2688667&-29.8246056 & 16.76 &  0.41 &2987 & 7183  & 6.8 & 4.1 \\
M83-9 & 204.2904083&-29.8187833 & 15.65 &  0.56 &1056 & 3617  & 3.6  & 2.5\\
M83-10 & 204.2746000&-29.8959056 & 16.94 &  0.34 &1780 & 5835  & 8.0  & 5.5 \\
M83-11 & 204.2179875&-29.8557389 & 16.65 & 0.35 & 2092 & 5653  & 5.4 & 3.2 \\
M83-12 & 204.2171125&-29.8764111 & 16.84 & 0.36 &2892 & 7277  & 6.3  & 4.3\\
M83-13 & 204.2127208&-29.8444639 & 16.76 & 0.43 & 2232 & 5653  & 5.6  & 3.4\\
M83-14 & 204.2230583&-29.8863750 & 15.30 & 0.35 &400 & 1296  & 2.8 & 1.7 \\
M83-15 & 204.2465708&-29.9075222 & 16.24 & 0.40 &1040 & 3644  & 2.1 & 1.6 \\
M83-16 & 204.2204833&-29.8887472 & 16.79 & 0.38 &2988 & 7169  & 5.4 & 3.4\\
M83-POS1 & 204.2519088 &-29.8651611 & 16.15 & 0.02 & 4093 & 1240 & 20.7 & 3.3\\
M83-POS2 & 204.2521171&-29.8670056 & 15.52 & 0.02& 2284& 420 & 30.6 & 2.9\\
\hline
\end{tabular}
\begin{minipage}{15cm}~\\
\textsuperscript{$\dagger$}{Coordinates extracted from the Mikulski Archive at the Space Telescope Science Institute (MAST). We note that the HST performance has jitter of 0.008$\arcsec\:$ RMS\footnote[4]{\href{https://hst-docs.stsci.edu/hsp/the-hubble-space-telescope-primer-for-cycle-28/hst-cycle-28-primer-optical-performance-guiding-performance-and-observing-efficiency}{https://hst-docs.stsci.edu/hsp}}.}\\
 \textsuperscript{$a$}{Magnitudes are in the Vegamag system, calculated using a 2.5$\arcsec\:$ aperture size.}\\
  \textsuperscript{$b$}{Calculated adopting the parameters listed in Table \ref{table:gal}.}\\
  \textsuperscript{$c$}{Estimated at wavelengths of 1310 \r{A} and 1700 \r{A} for G130M and G160M, respectively.}\\
 \end{minipage}
\end{table*}

\subsection{MODS observations}
Optical spectra of the \ion{H}{2} regions in M83 were acquired with MODS 
\citep[][]{pog10} on the LBT on the UT date of 2018 May 21.
The primary goal was to obtain high signal-to-noise spectra, with detections of the intrinsically 
faint auroral lines (e.g., [\ion{O}{3}] $\lambda$4363, [\ion{N}{2}] $\lambda$5755, [\ion{S}{3}] $\lambda$6312), in order to obtain accurate 
abundances of the gas surrounding the ionizing young massive clusters (YMCs) observed with COS and presented by \citet{her19}.
To do so, we used the multi-object mode of MODS, which uses custom-designed, laser-milled slit masks, 
allowing multiple \ion{H}{2} regions to be targeted simultaneously. We highlight that two masks were originally cut to cover the whole COS sample. However, due to poor weather conditions and other complications, only half of the data were collected. The M83 mask used, which targets 13 \ion{H}{2} regions simultaneously, was observed for three exposures of 1200s, 
or a total integration time of 1-hour.
At the latitude of the LBT, M83 stays below 30$^\circ$ on the sky, and thus the observations were obtained at 
relatively high airmass ($\sim$2--3).
To compensate, slits were cut close to the median parallactic angle of the observing window (PA=0), 
minimizing flux lost due to differential atmospheric refraction between 3200 -- 10,000 \r{A} \citep{fil82}.
Blue and red spectra were obtained simultaneously using the G400L (400 lines mm$^{-1}$, $R\sim1850$) 
and G670L (250 lines mm$^{-1}$ $R\sim2300$) gratings, respectively. 
The resulting combined spectra extend from 3200--10000 \AA, with a resolution of $\sim2$ \AA  (FWHM).

Broad R-band and continuum-subtracted H$\alpha$ images of M83 from The Spitzer Local Volume Legacy Survey 
\citep[LVL;][]{dale09} were used to identify the sample of target \ion{H}{2} regions, as well as the alignment stars.
\ion{H}{2} regions were selected by prioritizing the knots of high H$\alpha$ surface brightness that are
in the closest physical proximity to the YMCs.
All target slits are 1.0\arcsec$\:$ wide, but with lengths varying according to the size of each \ion{H}{2} region. 
The resulting multi-slit mask contained 13 \ion{H}{2} region slits and 2 sky slits. We list in the Appendix in Table \ref{tab:mods} the coordinates of each of the slits.
The mask slit locations are shown in Figure~\ref{fig:m83} in comparison to the stellar clusters.
Within the slit mask footprint, and avoiding aberration issues near the edges, we were able to target 
six distinct \ion{H}{2} regions that directly correspond to YMC regions. 
To effectively use the mask real estate, we targeted six additional \ion{H}{2} regions that do not clearly
correspond to one the of YMCs (R4, R5, R9, R13, R14, and R15).

The M83 optical spectra were reduced and analyzed using the MODS reduction pipeline\footnote{
\href{http://www.astronomy.ohio-state.edu/MODS/Software/modsIDL/}{http://www.astronomy.ohio-state.edu/MODS/Software/modsIDL/}} following the procedures detailed in \citet{berg15}.
Here, we summarize notable reduction steps.
Given the crowding of bright \ion{H}{2} regions in the disk of M83, diffuse nebular emission can complicate local sky subtraction. 
Therefore, the additional sky slits cut in the mask were used to provide a basis for clean sky subtraction. 
Continuum subtraction was performed in each slit by scaling the continuum flux from the sky-slit to the local background continuum level. 
One-dimensional spectra were then corrected for atmospheric extinction and flux calibrated based on observations 
of flux standard stars \citep{boh14}. 

\subsection{MUSE observations}
To complement our COS and MODS observations we make use of archival MUSE data covering 12 out of 17 clusters in our COS sample with  spectral resolution increasing between $R\sim$ 1770 at the bluest wavelengths (4800 \r{A}) and $R\sim$ 3590 at the reddest wavelengths (9300 \r{A}). The observations were taken during 2016 April and 2017 April--May as part of PID:  096.B-0057(A) with PI: Adamo. We downloaded the fully reduced datacubes from the European Southern Observatory (ESO) archive. The datacubes are reduced using the MUSE pipeline v.2.0.1, which performs removal of instrumental artifacts, astrometric calibration, sky-subtraction, wavelength and flux calibrations. We note that upon further inspection we found that the astrometric calibration for these archival datacubes was not correct, so we manually corrected their astrometry through a comparison with HST images. \par
We identify the location of the 12 clusters in our COS sample within the MUSE cubes and extract their spectra. The extraction was done by summing all spectra within circular apertures of diameter 2.5$\arcsec$ co-spatial with the HST/COS pointings, taking into account the fractional level of overlap of the spatial pixel with the region. We identified a few bad pixels in the datacubes and excluded them from our analysis. In the left panel of Figure \ref{fig:m83} we highlight in yellow the fields of view of MUSE.

\section{Analysis}\label{sec:ana}
\subsection{Neutral-gas Metallicities}
\subsubsection{Continuum and line-profile fitting}
As part of the analysis performed we normalized the individual COS spectroscopic observations before fitting the different line profiles. We fit the continuum of the star clusters by interpolating between regions (nodes) strategically positioned to avoid stellar and ISM absorption. We make use of a spline function when interpolating between the manually defined nodes. \par
We derive the column densities for the different elements by fitting Voigt profiles using the recently developed Python software \texttt{VoigtFit} v.0.10.3.3 \citep{kro18}. This relatively new code allows users to provide line spread function (LSF) tables to account for the broadening of the absorption lines introduced by the instrument itself. We convolve the COS LSF profiles with the FWHM of the source in the dispersion direction as measured from the acquisition images, similar to the approach described in Section 3.3 in \citet{her20}. We also note that \texttt{VoigtFit} allows for multi-component fitting, particularly useful for deblending different components along the same line of sight. \par
\begin{figure*}
   \centering
   	  \centerline{\includegraphics[scale=0.5]{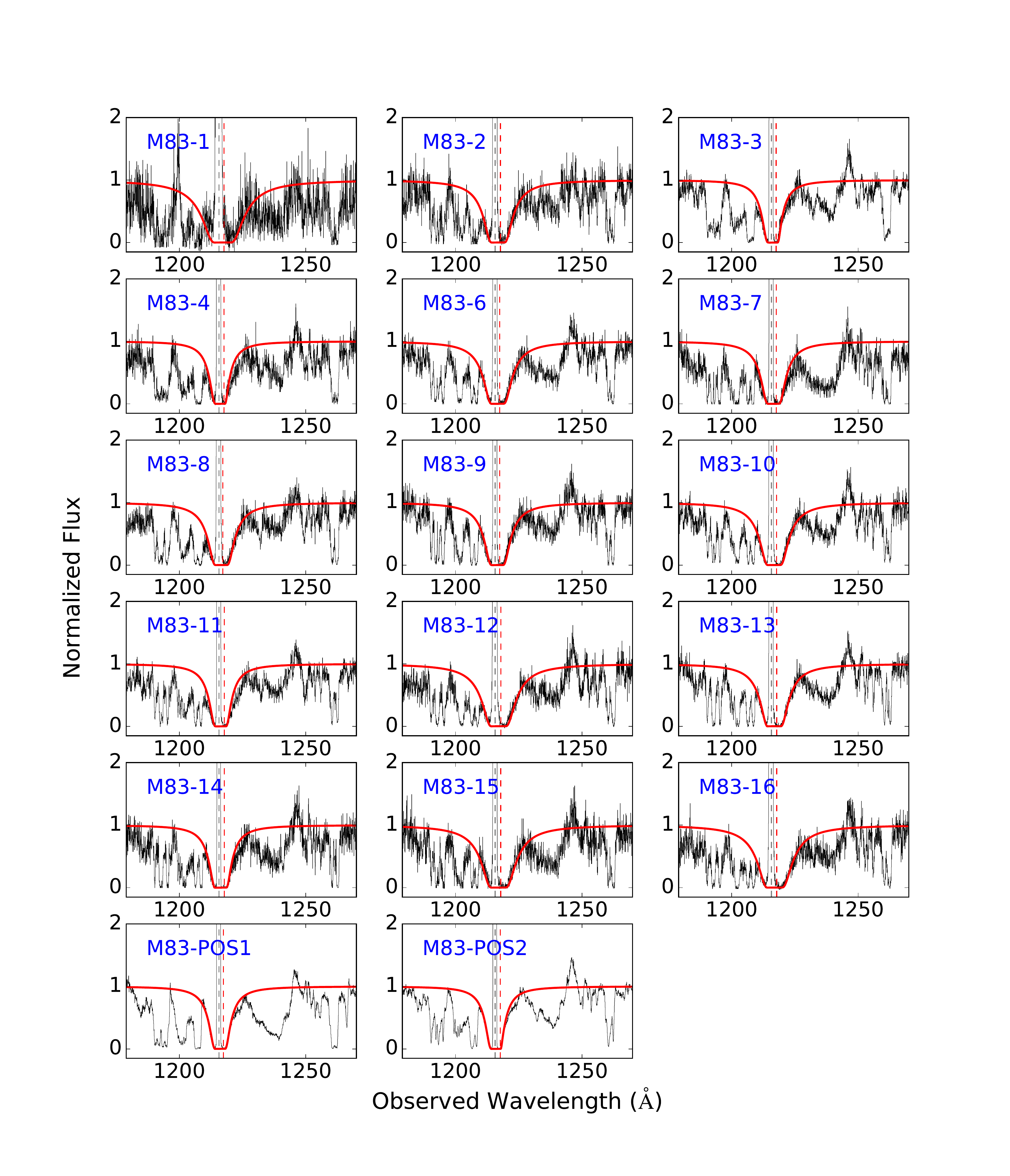}}
      \caption{Ly$\alpha$ profiles for the M83 pointings in our sample. In black we show the COS observations binned by 1 resolution element (1 resel = 6 pixels). In red we display the best fitting VoigtFit model. The names of the individual targets are shown in each panel. We show with thin dashed lines the MW component (grey) and the M83 component (red). }
         \label{fig:HI}
   \end{figure*}
\begin{figure*}
   	  \centerline{\includegraphics[scale=0.5]{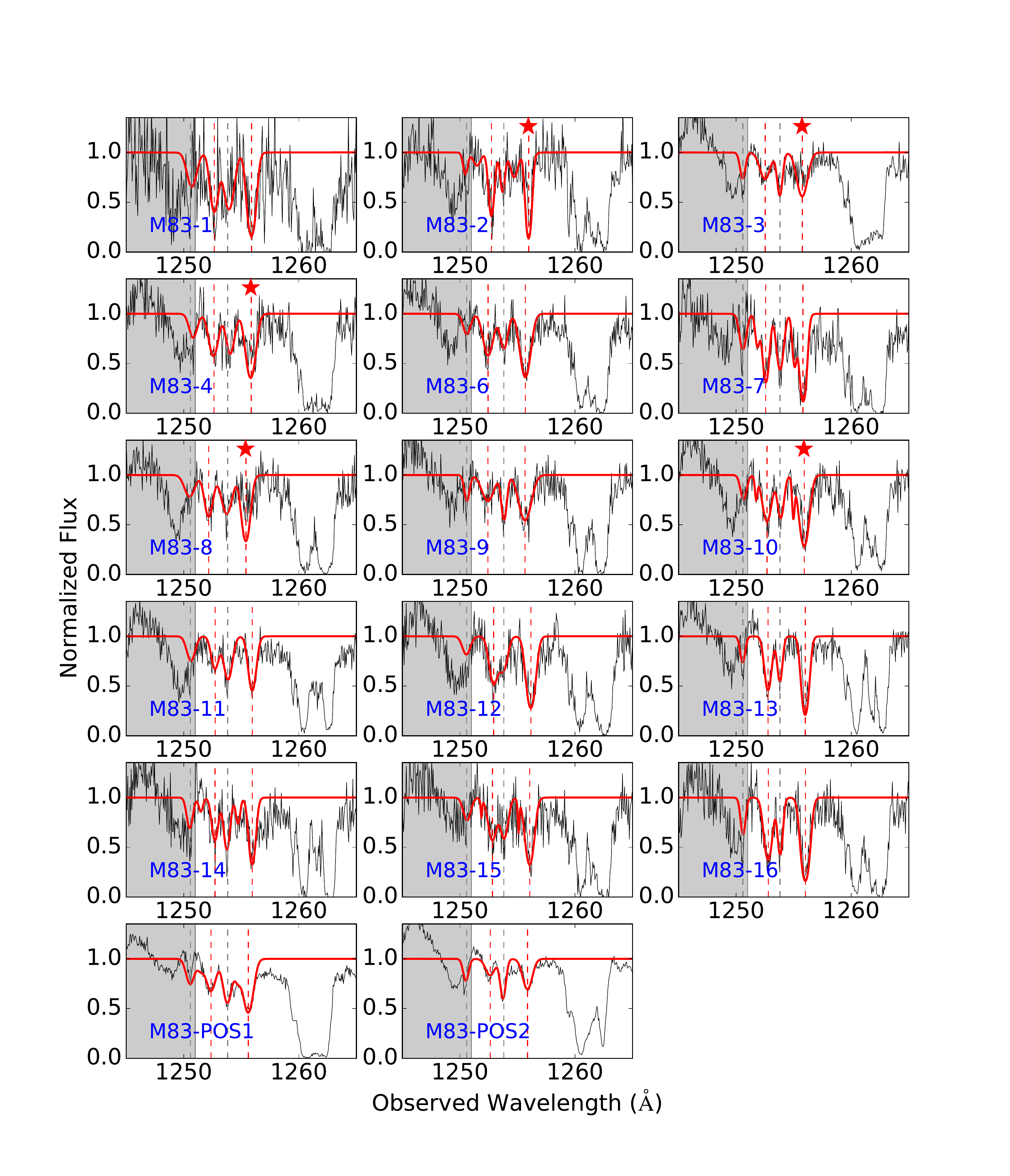}}
      \caption{\ion{S}{2} profiles for the M83 pointings in our sample. In black we show the COS observations binned by 1 resolution element (1 resel = 6 pixels). In red we display the best fitting model. The names of the individual targets are shown in each panel. Vertical grey dashed lines show the location of the MW components, vertical red dashed lines indicate the strongest M83 components. We note that we have masked out the MW \ion{S}{2} $\lambda$1250 line when fitting our extragalactic \ion{S}{2} lines, as this is strongly affected by the P-Cygni profile of the \ion{N}{5} line. We show these masks as shaded grey regions. Lastly, we mark those pointings exhibiting hidden saturation with a red $\bigstar$. The fits for these targets have been obtained excluding the strongest \ion{S}{2} $\lambda$1253 line.}
         \label{fig:SII}
   \end{figure*}

We note that although the COS observations allow us to access a variety of absorption lines of several heavy elements, our work focuses on measuring the metallicity of the neutral gas. Given that S/H traces metallicity reliably \citep{leb13,jam18} we primarily study the \ion{S}{2} and Ly$\alpha$ lines. We present in Table \ref{tab:line_pars} the theoretical parameters for each of the lines analyzed as part of this work. We show in Figures \ref{fig:HI} and \ref{fig:SII} the best fitting profiles for the Lyman $\alpha$ and \ion{S}{2} lines along with the COS observations. 

\begin{table}
\caption{Atomic Data for UV absorption lines}
\label{tab:line_pars}
\centering 
\begin{tabular}{ccc}
\hline \hline
Line ID & $\lambda_{\rm rest}$ & $f^{a}$ \\
& (\AA) &\\
\hline\\
Ly$\alpha$ & 1215.6710 & 4.16e-01\\
\ion{S}{2} &1250.5780 &5.43e-03 \\
\ion{S}{2} & 1253.8050& 1.09e-02\\
\hline
\end{tabular}
\begin{minipage}{15cm}~\\
 \textsuperscript{$a$}{Oscillator strength values compiled by the Vienna Atomic \\
 Line Database 3 (VALD3).}\\
 \end{minipage}
\end{table}

\subsubsection{\ion{H}{1}}\label{sec:HI}
The COS observations analyzed here cover the Ly$\alpha$ absorption line at $\lambda = $1215.671 \AA$\:$  originating from the multiple sightlines in M83. Given the close proximity of M83, the Ly$\alpha$ absorption from the MW is heavily blended with those from our M83 pointings. In order to extract precise column densities for the \ion{H}{1} gas in M83 we simultaneously fit the MW and galaxy Ly$\alpha$ profiles. Following the approach adopted by \citet{jam14}, we make use of the red wing of Ly$\alpha$ to constrain the fit of the \ion{H}{1} column density intrinsic to the different targets,  and adopt a fixed MW \ion{H}{1} column density measured in \citet{jam14} in the direction of M83, $ \log$[$N$(\ion{H}{1})$_{\rm MW}$ cm$^{-2}$] = 20.57. Measurements of the \ion{H}{1} column densities of the individual pointings in M83, $\log[$$N$(\ion{H}{1})], are listed in Table \ref{tab:HI}. Lastly, the best fitting profiles for the whole sample are shown in different panels in Figure \ref{fig:HI}.

\subsubsection{\ion{S}{2}}\label{sec:SII}
In general, direct measurements of the oxygen abundances in the neutral gas are difficult to access as the most easily observed \ion{O}{1} line, 1302 \r{A}, in the COS spectral coverage at low redshifts is typically saturated. On the other hand, \ion{O}{1} at 1355 \r{A} is too weak to be detected. As such, we use proxies for oxygen to indirectly derive the oxygen abundances \citep{jam18}. As part of our analysis we measure the column densities of the \ion{S}{2} lines listed in Table \ref{tab:line_pars} and make use of the solar ratio of $\log$(S/O)$_{\odot}$ = $-$1.57 $\pm$ 0.06 to derive the O/H abundances in the neutral gas of M83. 
   
\subsubsection{Curve-of-growth Analysis}\label{sec:COG}   
To assess if our measured abundances are affected by saturation, we plot the column density measurements along the curve-of-growth (COG) corresponding to the ion in question. For each ion we generate a COG showing the relation between the equivalent width, $\log$($W/\lambda$), and the column density, $\log$($fN$), where $f$ is the oscillator strength. When generating the individual COGs we adopt the $b$ parameters listed in Appendix \ref{app:bval} in Table \ref{tab:bval} inferred from the simultaneous line-profile fitting of the \ion{S}{2}
 lines. \par
\begin{figure*}
   	  \centerline{\includegraphics[scale=1.45]{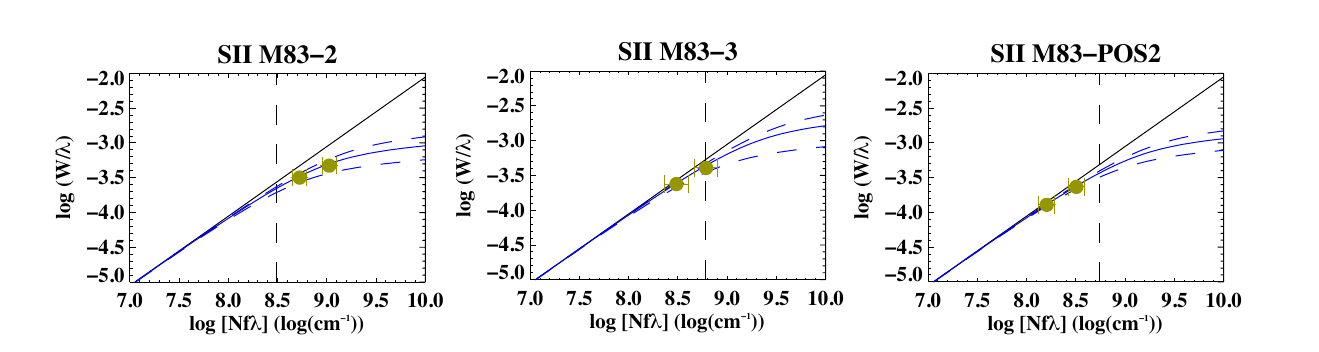}}
      \caption{Selected sample of curves of growth displaying the linear and saturated regimes for a fitted $b$ parameter specific  to \ion{S}{2}. We show with blue dashed lines the 1$\sigma$ errors on the $b$ parameter. We indicated with a dashed vertical line the transition from the linear to the saturated regime. Each subplot illustrates a different line-strength regime. The filled circles show the equivalent width, $W$, and column density, $N$, of each line as derived from our line-profile fitting analysis. Left: \ion{S}{2} transitions in the saturated part of the COG indicating saturation in both lines. Middle: One of the two transitions lies close to the saturated regime. The location of the second transition in the linear regime allows us to rule out the possibility of hidden saturation e.g., if we are able to fit both lines simultaneously. Right: Both transitions show unsaturated lines. }
         \label{fig:COG}
   \end{figure*}

 In Figure \ref{fig:COG} we show a selected sample of COGs illustrating the line strength regimes encountered in our analysis. We primarily use the location of the \ion{S}{2} transitions on their corresponding COG to determine if the column density estimates are reliable or if they need to be considered as lower limits due to saturation effects. For those pointings where both \ion{S}{2} transitions are found on the right side of the vertical line, we consider them as saturated lines as they occupy the curved or saturated regime in the COG (see left panel in Figure \ref{fig:COG}). In those cases we are unable to constrain the column densities for \ion{S}{2} and we consider them as lower limits. We also identified cases where one of the two transitions was found borderline or clearly in the saturated regime as shown in the middle panel of Figure \ref{fig:COG}. In such scenario we are still able to constrain the column densities as we fit the weakest transition, i.e., \ion{S}{2} $\lambda$1250, avoiding hidden saturation effects. Lastly, we observed cases where both transitions were located on the linear part of the COG clearly showing an absence of saturation for those pointings. 
 
\begin{table*}
\caption{Column densities for the different pointings in M83. }
\label{tab:HI}
\centering 
\begin{tabular}{ccccc|cc}
\hline \hline
Target & $\log$[$N$(\ion{H}{1})] & $\log$[$N$(\ion{S}{2})]& $\log$[$N$(\ion{S}{2})]$_{2}$$^{a}$& $\log$[$N$(\ion{S}{2})]$_{\rm TOTAL}$ & $\log$[$N$(\ion{H}{1})]$_{\rm ICF}$$^{b}$ &  $\log$[$N$(\ion{S}{2})]$_{\rm TOTAL\_ICF}$$^{b}$\\
 & (cm$^{-2}$)  & (cm$^{-2}$)  & (cm$^{-2}$)  & (cm$^{-2}$) & (cm$^{-2}$) &  (cm$^{-2}$) \\
\hline\\
M83-1 &  21.05 $\pm$ 0.08 & 15.99 $\pm$ 0.07$^\dagger$ &- &$>$15.99$^\dagger$ & 21.17 $\pm$ 0.08 & $>$16.09$^\dagger$ \\
M83-2 &20.71 $\pm$ 0.08 &15.89 $\pm$ 0.07$^\dagger$ &15.17 $\pm$ 0.11 &$>$15.97$^\dagger$ & 20.82 $\pm$ 0.08 & $>$16.05$^\dagger$\\
M83-3 & 18.99 $\pm$ 0.23 & 15.65 $\pm$0.12 & -&15.65 $\pm$0.12 & 18.94 $\pm$ 0.23 & 15.25 $\pm$ 0.12\\
M83-4 & 19.56 $\pm$ 0.21 &15.85 $\pm$ 0.08 &-&15.85 $\pm$ 0.08 & 19.55 $\pm$ 0.21 & 15.68 $\pm$ 0.08 \\
M83-5 & 21.01 $\pm$ 0.18 & -&- & - & - & -\\
M83-6 & 20.60 $\pm$ 0.06& 15.88 $\pm$ 0.07&- & 15.88 $\pm$ 0.07 & 20.72 $\pm$ 0.06 & 15.96 $\pm$ 0.07\\
M83-7 & 20.65 $\pm$ 0.09& 16.00 $\pm$ 0.05$^\dagger$ & 15.38 $\pm$ 0.13 & $>$16.09$^\dagger$ & 20.65 $\pm$ 0.09 & $>$16.06$^\dagger$  \\
M83-8 & 20.54 $\pm$ 0.03 & 15.79 $\pm$ 0.06 & -& 15.79 $\pm$ 0.06 & 20.65 $\pm$ 0.03 & 15.86 $\pm$ 0.06 \\
M83-9 & 20.45 $\pm$ 0.07 & 15.70 $\pm$ 0.04 & -& 15.70 $\pm$ 0.04 & 20.56 $\pm$ 0.07 & 15.77 $\pm$ 0.04\\
M83-10 & 20.62 $\pm$ 0.02 & 15.88 $\pm$ 0.05$^\dagger$ & 15.01 $\pm$ 0.22& $>$15.94$^\dagger$& 20.74 $\pm$ 0.02 & $>$16.02$^\dagger$  \\
M83-11 & 20.05 $\pm$ 0.07 & 15.61 $\pm$ 0.04 & -&15.61 $\pm$ 0.04 & 20.20 $\pm$ 0.07 & 15.68 $\pm$ 0.04 \\
M83-12 &20.85 $\pm$ 0.02 & 15.88 $\pm$ 0.03 &- & 15.88 $\pm$ 0.03 & 20.98 $\pm$ 0.02 & 15.97 $\pm$ 0.03 \\
M83-13 & 20.63 $\pm$ 0.02 & 15.87 $\pm$ 0.05$^\dagger$ &- & $>$15.87$^\dagger$ & 20.75 $\pm$ 0.02 & $>$15.95$^\dagger$  \\
M83-14 & 20.43 $\pm$ 0.13 &15.70 $\pm$ 0.05 & 14.96 $\pm$ 0.15& 15.77 $\pm$ 0.05 & 20.43 $\pm$ 0.13 & 15.73 $\pm$ 0.05\\
M83-15 & 20.91 $\pm$ 0.05 & 15.81 $\pm$ 0.09 & 14.72 $\pm$ 0.47& 15.84 $\pm$ 0.09 & 21.04 $\pm$ 0.05 & 15.94 $\pm$ 0.09\\
M83-16 & 21.05 $\pm$ 0.02 & 16.04 $\pm$ 0.04$^\dagger$ & -& $>$16.04$^\dagger$ & 21.18 $\pm$ 0.02 & $>$16.15$^\dagger$\\
M83-POS1 & 19.93 $\pm$ 0.03 & 15.71 $\pm$ 0.06 & 15.02 $\pm$ 0.21 & 15.79 $\pm$ 0.06 & 19.92 $\pm$ 0.03 & 15.63 $\pm$  0.06\\
M83-POS2 & 19.02 $\pm$ 0.03 & 15.37 $\pm$ 0.08 & -& 15.37 $\pm$ 0.08 & 18.91 $\pm$ 0.03 & 14.72 $\pm$ 0.08 \\
\hline
\end{tabular}
\begin{minipage}{15cm}~\\
 \textsuperscript{$a$}{Multi-component cases. A second \ion{S}{2} component was identified for these pointings.}\\
  \textsuperscript{$b$}{Column densities calculated after applying the ionization correction factors listed in Table \ref{tab:icf_ionized}.}\\
  \textsuperscript{$^\dagger$}{Column densities should be considered as lower limits.}\\
 \end{minipage}
\end{table*}

\subsubsection{Ionization corrections}
In ISM abundance studies it is critical to take into account ionization effects due to contaminating ionized gas along the line of sight and/or contributing higher ionization ions present in the neutral gas, but not measured directly from the observations. Generating tailored photoionization models for a sample of nearby star-forming galaxies, \citet[][hereafter H20]{her20} found ionization correction factors (ICFs) as high as $\sim$0.7 dex in the neutral gas, clearly demonstrating the importance of precise ICFs. To accurately infer the chemical abundances of the neutral gas along the different pointings throughout M83, as part of this work we investigate the amount of ionized gas contaminating the neutral abundance measurements (ICF$_{\rm ionized}$), as well as the amount of higher ionization ions, compared to the dominant ion of a certain species in the \ion{H}{1} gas (ICF$_{\rm neutral}$). \par
To accurately estimate the ionization effects affecting our measured abundances we adopt a similar approach as that followed by H20. We generate tailored photoionization models for each of the pointings in our sample using the spectral synthesis code \texttt{CLOUDY} \citep{fer17}. We adopt an overall metallicity of $Z$= 3.24 $Z_{\odot}$ as measured from the ionized-gas component \citep{mar10}. The work of H20 takes advantage of newly acquired COS/FUV observations covering bluer wavelengths than our M83 COS/FUV data, which they use to measure the \ion{Fe}{3}/\ion{Fe}{2} ratio for the galaxies in their sample, including two M83 pointings in our analysis here (M83-POS1 and M83-POS2). This ratio is a critical indicator of the gas volume density of the targets and is essential to generate tailored photoionization models. Given that the \ion{Fe}{3}  line at $\lambda=$ 1122 $\rm \AA\;$ is not covered in our COS observations, we instead adopt an average value from the two M83 pointings in H20 of  $\log$[$N$(\ion{Fe}{3})/$N$(\ion{Fe}{2})] = $-$0.811 dex for the rest of the M83 targets studied here. Furthermore, H20 estimate the effective temperature of the star clusters observed in the two M83 pointings in their sample to be  T$_{\rm eff}=$42,500 K. We adopt the same T$_{\rm eff}$ for the rest of our M83 pointings. We highlight that according to the work by H20, this physical parameter (T$_{\rm eff}$) has minimal effects on the final ICF values calculated from the photoionization models. The rest of the input parameters are listed in Table \ref{tab:cloudy}. We estimate the log[L$_{\rm UV}$] values from the $m_{\rm F336W}$ values listed in the Table \ref{table:obs}, with the exception of M83-POS1 and M83-POS2; the log[L$_{\rm UV}$] values for these two clusters were calculated using ACS/SBC frames observed with the F125LP filter. We measure the \ion{H}{1} column densities directly from the COS observations as described in section \ref{sec:HI}. In the third column of Table \ref{tab:cloudy} we show the measured volume densities using the assumed \ion{Fe}{3}/\ion{Fe}{2} ratios. 
For more details on the precise steps taken to generate the photoionzation models we refer the reader to \citet{her20}. \par
The different ICF values, ICF$_{\rm ionized}$, ICF$_{\rm neutral}$, and ICF$_{\rm TOTAL}$, for the full M83 sample are listed on Table \ref{tab:icf_ionized}. Similar to the work of  \citet{jam14} and H20 we calculate the final column densities for each element X using the following equation:
\begin{equation}\label{eq:corr_col}
\log[N(X)_{\rm ICF}] = \log[N(X)] - \rm ICF_{\rm TOTAL}
\end{equation}
In Table \ref{tab:icf_ionized} we list the individual correction values, both ICF$_{\rm ionized}$ and ICF$_{\rm neutral}$, for each ion and target. In the last two columns of Table \ref{tab:icf_ionized} we show the total ionization correction factors to be applied to the measured column densities. And finally, we list in the last columns of Table \ref{tab:HI} the ionization corrected column densities for H and S obtained after applying the inferred ICF$_{\rm TOTAL}$ using Equation \ref{eq:corr_col}. 

\begin{table}
\caption{Input parameters for the \texttt{CLOUDY} models tailored to each of the M83 pointings in our sample}
\label{tab:cloudy}
\centering 
\begin{tabular}{ccc}
\hline \hline
Target & $\log$[L$_{\rm UV}$]$^{a}$ & log[$n$(H)]\\
  & (erg s$^{-1}$) &  (cm$^{-3}$) \\
\hline\\
        M83-1& 38.25 & 1.14 \\	 	 
	M83-2&38.60 & 2.31\\ 	 	 	  	 
	M83-3&39.01 & 3.85\\	 	 	  	  	 	 
	M83-4&39.30 & 3.97\\ 	
	M83-5&39.12 & 2.53 \\
	M83-6&38.48 & 2.29\\ 	  	 
	M83-7& 39.14& 3.06\\
	M83-8&38.36 & 2.22 \\	 	 	 		 	  
	M83-9& 38.80 & 2.86\\ 
	M83-10&38.28 & 2.00\\ 
	M83-11& 38.40 & 2.72\\ 
	M83-12&38.32& 1.70\\ 
	M83-13&38.36 & 2.09\\ 
	M83-14&38.94 & 3.05\\ 
	M83-15&38.56 & 1.93\\ 
	M83-16&38.34 &1.20\\ 		 	 	 	 	 
	M83-POS1& 40.94$^b$& 4.68$^{c}$\\	 	 	 	 	 
	M83-POS2 & 41.40$^b$& 5.72$^{c}$\\	 	 	 	 	 
\hline
\end{tabular}
\begin{minipage}{15cm}~\\
\textsuperscript{$a$}{Luminosities estimated from the $m_{\rm F336W}$ listed in Table \ref{table:obs}.}\\
\textsuperscript{$b$}{Luminosities estimated from the ACS/SBC frames \\
observed with the F125LP filter \citep{her20}.}\\
\textsuperscript{$c$}{Volume densities from \citet{her20}.}\\
 \end{minipage}
\end{table}

\begin{table}
\caption{Ionization Correction Factors for the M83 COS pointings}
\label{tab:icf_ionized}
\centering 
\begin{tabular}{cccccrr}
\hline \hline
& \multicolumn{2}{c}{ICF$_{\rm ionized}$}&\multicolumn{2}{c}{ICF$_{\rm neutral}$} &\multicolumn{2}{c}{ICF$_{\rm TOTAL}$}\\
 \hline
Target & \ion{H}{1} &  \ion{S}{2} & \ion{H}{2} &  \ion{S}{3} &\multicolumn{1}{c}{H}  &   \multicolumn{1}{c}{S}\\
\hline\
        M83-1& 0.00 &   0.00 & 0.12 &   0.10& $-$0.12 &  $-$0.10 \\	 	 
	M83-2 & 0.00 &   0.00 & 0.12 &   0.08 & $-$0.12 &   $-$0.08\\ 	 
	M83-3& 0.08 &   0.40 & 0.04 &   0.01  & 0.04 &   0.40 \\	
	M83-4& 0.02 &   0.17 & 0.01 &   0.00& 0.01 &   0.17\\ 	
	M83-5& 0.00 &   0.00 & 0.12 &   0.10 & $-$0.12 &   $-$0.10\\
	M83-6 & 0.00 &   0.00 & 0.12 &   0.08 & $-$0.12 &   $-$0.08 \\ 	  	 
	M83-7& 0.00 &   0.03  & 0.00 &   0.00  & $-$0.00 &   0.03   \\
	M83-8& 0.00 &   0.00 & 0.12 &   0.08 & $-$0.12 &  $-$0.08 \\	 
	M83-9& 0.00 &   0.00 & 0.11 &   0.07 & $-$0.11 &  $-$0.07  \\ 	 	
	M83-10& 0.00 &   0.00 & 0.12 &   0.08  & $-$0.12 &   $-$0.08 \\ 
	M83-11& 0.00 &   0.00  & 0.15 &   0.07 & $-$0.15 &   $-$0.07\\ 
	M83-12& 0.01 &   0.00  & 0.12 &   0.10& $-$0.12 &   $-$0.10 \\ 
	M83-13& 0.00 &   0.00  & 0.12 &   0.08 & $-$0.12 &   $-$0.08\\ 
	M83-14& 0.00 &  0.04  & 0.01 &   0.00 & $-$0.00 &   0.04 \\ 
	M83-15& 0.00 &   0.00  & 0.12 &   0.10 & $-$0.12 &   $-$0.10 \\ 
	M83-16& 0.00 &   0.00 & 0.13 &   0.11 & $-$0.13 &   $-$0.11  \\  
	M83-POS1& 0.02 &   0.16 & 0.01 &   0.00 &   0.01 &  0.15 \\	 	 
	M83-POS2& 0.27 &   0.67& 0.17 &   0.02& 0.11 &  0.65  \\	 	 	 	 	 
 \hline
 \end{tabular}
\end{table}
 
\subsection{Nebular Metallicities}
We measure the emission line fluxes from the optical observations (LBT and VLT) for the recombination and collisionally excited lines by fitting Gaussian profiles after subtracting the continuum and absorption features in the spectral region of interest. Equal weight is given to the flux in each spectral pixel while fitting the Gaussian profiles. We further propagate the uncertainties on the three Gaussian parameters (amplitude, centroid, and FWHM) to estimate the final uncertainty in the fluxes. \par
We use the attenuation curve by \citet{fit19} along with the observed H$\alpha$/H$\beta$ ratio to estimate the nebular emission line color excess, E(B-V), at an electron temperature and density of 10,000 K and 100 cm$^{-3}$, respectively (Case B recombination). We note that we have also tested our metallicity calculations assuming Case B recombination coefficients associated with an electron temperature of 5000 K (which may be more representative of the high metallicity gas within M83), and find that the final metallicity estimates are insensitive to the electron temperature adopted for the Case B recombination coefficient, within the uncertainties. The E(B-V) is then used to deredden the observed emission line fluxes. We include in Appendix \ref{app:optical} the tables listing the individual fluxes, dereddened fluxes and reddening values for each of the pointings studied here. \par
As part of our analysis we tested various diagnostics for estimating the gas-phase metallicities, which included R$_{23}$\footnote{([\ion{O}{2}] $\lambda$3727+[\ion{O}{3}] $\lambda\lambda$4959, 5007)/H$\beta$}, O3N2\footnote{([\ion{O}{3}] $\lambda$5007/H$\beta$)/([\ion{N}{2}] $\lambda$6584/H$\alpha$)}, and N2\footnote{[\ion{N}{2}] $\lambda$6584/H$\alpha$} \citep{pet04, cur17}. We note that the \citet{cur17} metallicity calibrations are only valid for 12+$\log$(O/H) $<$ 8.85, objects with metallicities of 12+$\log$(O/H) $=$ 8.85 need to be considered lower limits. This limitation drastically reduced the number of available metallicity measurements in our study, as we are primarily exploring a high-metallicity environment. In a previous metallicity study of M83 by \citet{bre16}, they find that empirically calibrated strong-line diagnostics usually provide lower abundances than those inferred from the stellar populations. They attribute this behavior to the difficulties in selecting adequate samples when calibrating high metallicity environments. They note that among those strong-line methods tested in their work, the O3N2 calibration by \citet{pet04} provides nebular abundances that are in best agreement with their BSG metallicities. For the rest of our study we adopt the O3N2 calibration by \citet{pet04} as recommended by \citet{bre16}. The uncertainties in the final nebular metallicities listed in the last two columns of Table \ref{tab:z} account for both the statistical and systematic components. We highlight that pointings  M83-1 and M83-6 have been observed with both LBT and VLT. The metallicities calculated from these two sets of observations agree within their uncertainties.\par
Lastly, we performed a detailed inspection on possible contamination on our nebular fluxes due to nearby supernova remnants (SNRs). Using the catalogs by \citet{bla14}, \citet{bla12},  \citet{dop10} and \citet{rus20} of previously identified SNRs in M83, we confirm that almost all of the slits and apertures are free of contamination, with the exception of LBT target R7 (see Table \ref{tab:mods}). We exclude this target from the rest of our analysis. \par
 
 \subsection{Stellar Metallicities}
\citet{her19} performed the first metallicity study of M83 using the integrated UV light of most of the YMCs we study here. More precisely they measure the metallicities of those targets observed in HST PID 14681. \citet{her19} did not include the last two targets listed in Table \ref{table:obs} from HST PID:  11579 and 15193, M83-POS1 and M83-POS2. They applied the same full spectral fitting technique developed by \citet{lar12}, and previously applied to spectroscopic observations of stellar populations in the optical and NIR wavelength regime. Briefly summarized, this technique combines the information from the Hertzsprung-Russell diagram, stellar atmospheric models and synthetic spectra to derive abundances from the integrated light of single stellar populations. \par
In order to have stellar metallicities of our full M83 sample, we apply the same approach as that described in \citet{her19} to measure the overall metallicities of the two missing clusters, M83-POS1 and M83-POS2. After some inspection of the individual targets and their acquisition images, we discovered that the 2.5\arcsec$\:$ COS aperture for M83-POS2 encompassed more than one YMC. Given that the COS observations for this target contained multiple stellar population, we were unable to estimate the stellar metallicity for this pointing as the analysis technique by \citet{lar12} is optimized for single stellar populations. \par
We adopt an age of 3 Myr \citep{wof11} when fitting for the stellar metallicity of M83-POS1, and measure an overall metallicity of [Z] = $\log$Z/Z$_{\odot}$ = $+$0.18 $\pm$ 0.12 dex. We adopt the solar oxygen abundance by \citet{asp09}, 12+$\log$(O/H) = 8.69, and obtain an oxygen abundance of 12+$\log$(O/H) = 8.87 for M83-POS1. \par

We list the final metallicities from all three components in Table \ref{tab:z}. 

\begin{table*}
\caption{M83 metallicities of the stellar, neutral-gas, ionized-gas components for the COS pointings.}
\label{tab:z}
\centering 
\begin{tabular}{ccccc}
\hline \hline
& HST/COS & HST/COS & VLT/MUSE & LBT/MODS \\
Target & 12+log(O/H)$_{\rm stellar}$ & 12+log(O/H)$_{\rm neutral}$ & 12+log(O/H)$_{\rm ionized}$ & 12+log(O/H)$_{\rm ionized}$\\
& & & O3N2 &  O3N2 \\
\hline\
        M83-1&9.26 $\pm$ 0.10 &$>$8.48$^\dagger$ & 8.80 $\pm$ 0.15 & 8.97 $\pm$ 0.14 \\	 	 
	M83-2 & 8.55 $\pm$ 0.17 &$>$8.80$^\dagger$ & -- & --\\ 	 
	M83-3& 9.02 $\pm$ 0.15 &9.88 $\pm$ 0.27 & 8.85 $\pm$ 0.15 & --\\	
	M83-4& 8.71 $\pm$ 0.16&9.70 $\pm$ 0.23   & 8.86 $\pm$ 0.14  & -- \\ 	
	M83-5&-- &--  & --& --  \\
	M83-6 & 8.74 $\pm$ 0.12&8.81 $\pm$ 0.11 & 8.86 $\pm$ 0.20 & 9.00 $\pm$ 0.16 \\ 	  	 
	M83-7& 8.90 $\pm$ 0.18& $>$8.99$^\dagger$ & -- & --\\
	M83-8& 8.65 $\pm$ 0.14 &8.78 $\pm$ 0.09  & -- & 8.93 $\pm$ 0.14\\	 
	M83-9& 8.35 $\pm$ 0.08 &8.77 $\pm$ 0.10 &--  & 8.64 $\pm$ 0.14 \\ 	 	
	M83-10& 8.89 $\pm$ 0.15 & $>$8.85$^\dagger$& 8.84 $\pm$ 0.14 & -- \\ 
	M83-11& 8.66 $\pm$ 0.09 &9.05 $\pm$ 0.10 & -- & -- \\ 
	M83-12& 8.81 $\pm$ 0.14 &8.57 $\pm$ 0.07 & 8.84 $\pm$ 0.15 & --\\ 
	M83-13& 8.75 $\pm$ 0.13 & $>$8.77$^\dagger$ & -- & -- \\ 
	M83-14& 8.81 $\pm$ 0.19 &8.87 $\pm$ 0.15& 8.89 $\pm$ 0.14 & -- \\ 
	M83-15& 8.62 $\pm$ 0.08 &8.47 $\pm$ 0.12 & --  & --\\ 
	M83-16& 8.87 $\pm$ 0.14 & $>$8.53$^\dagger$   & 8.86 $\pm$ 0.14 & --\\  
	M83-POS1& 8.87 $\pm$ 0.12 &9.28 $\pm$ 0.09 & 9.00  $\pm$ 0.14  & -- \\	 	 
	M83-POS2&-- &9.38 $\pm$ 0.10 & 8.89 $\pm$ 0.14 & -- \\	 	 	 	 	 
 \hline
 \end{tabular}
 \begin{minipage}{15cm}~\\
  \textsuperscript{$^\dagger$}{Metallicities should be considered as lower limits.}\\
 \end{minipage}
\end{table*}

 \section{Discussion}\label{sec:discussion}
 \subsection{\ion{H}{1} distribution}\label{sec:HI}
\begin{figure}
   	  \centerline{\includegraphics[scale=0.38]{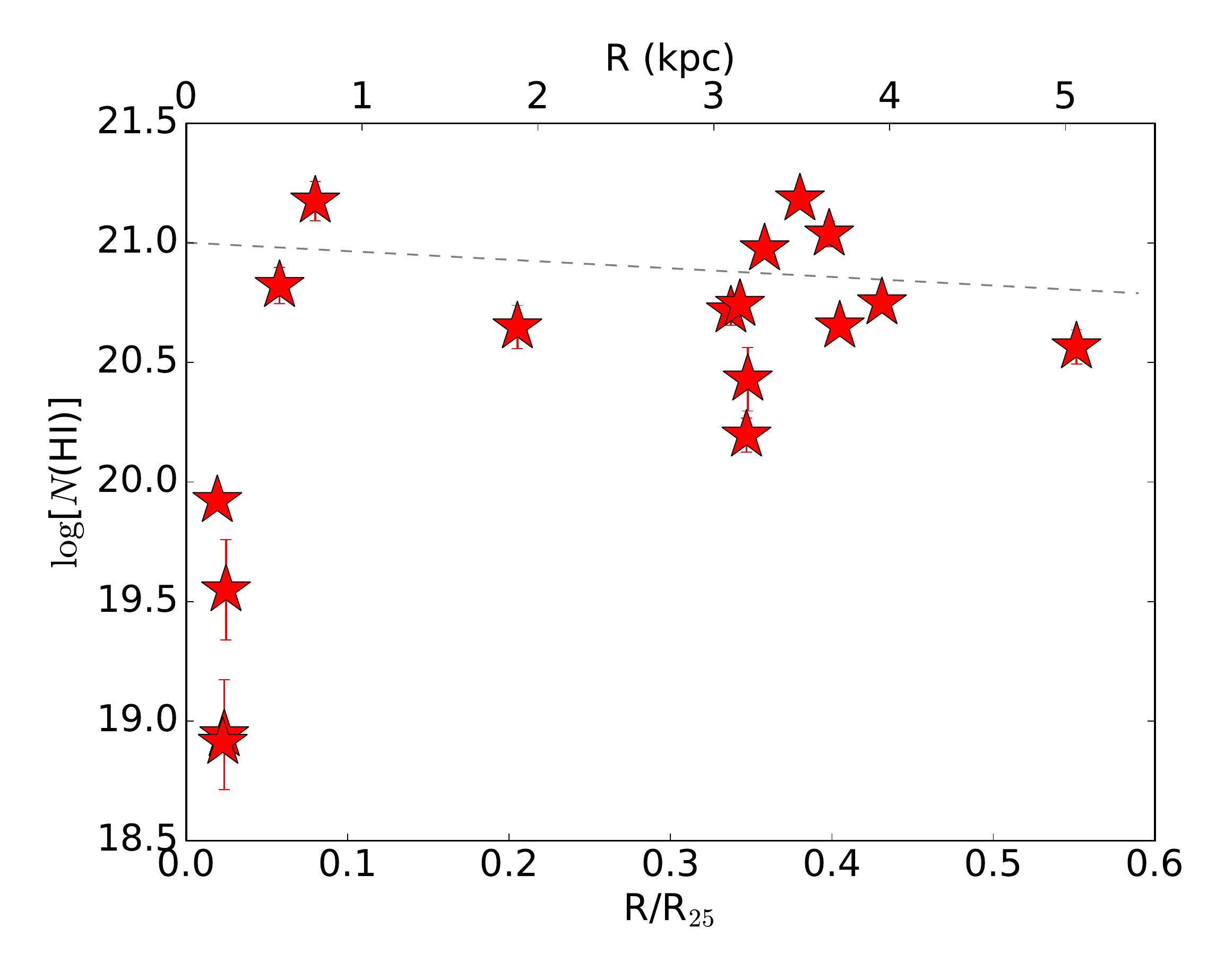}}
      \caption{\ion{H}{1} column densities measured from the COS observations as a function of galactocentric distance (bottom axis: normalized to isophotal radius). A linear regression is shown for those M83 pointings with $\log$[$N$(\ion{H}{1})]$>$ 20.0 dex. }
         \label{fig:HI_gradient}
   \end{figure}

 We show in Figure \ref{fig:HI_gradient} the \ion{H}{1} column density as a function of galactocentric distance, normalized to isophotal radius (see Table \ref{table:gal}). We find a depletion of \ion{H}{1} gas in the nuclear region of M83, with column densities of $\log$[$N$(\ion{H}{1}) cm$^{-2}$] $<$20.0. Our data indicate a general trend where at galactocentric distances R/R$_{\rm 25}>$0.02 the column density of \ion{H}{1} increases to values of the order of $\log$[$N$(\ion{H}{1}) cm$^{-2}$]$\sim$ 21.0, typical of the disks of spiral galaxies \citep{kam92,big08,ian18}, with a relatively flat gradient to larger galactocentric distances of $-$0.4 $\pm$ 1.1 dex $R_{\rm 25}^{-1}$ (dashed line in Figure \ref{fig:HI_gradient}). \par
 \citet{lun04} found that the CO emission in M83, which is assumed to be linearly proportional to the mass surface density intensity of the molecular hydrogen (H$_2$), and the \ion{H}{1} column density follow each other tightly with one clear exception at the nucleus, where they observed a clear depletion of \ion{H}{1}.  The low column densities reported in our study in the center of M83 clearly agree with the molecular- and neutral-gas maps in \citet{lun04}. \par
\begin{figure*}
   	  \centerline{\includegraphics[scale=0.51]{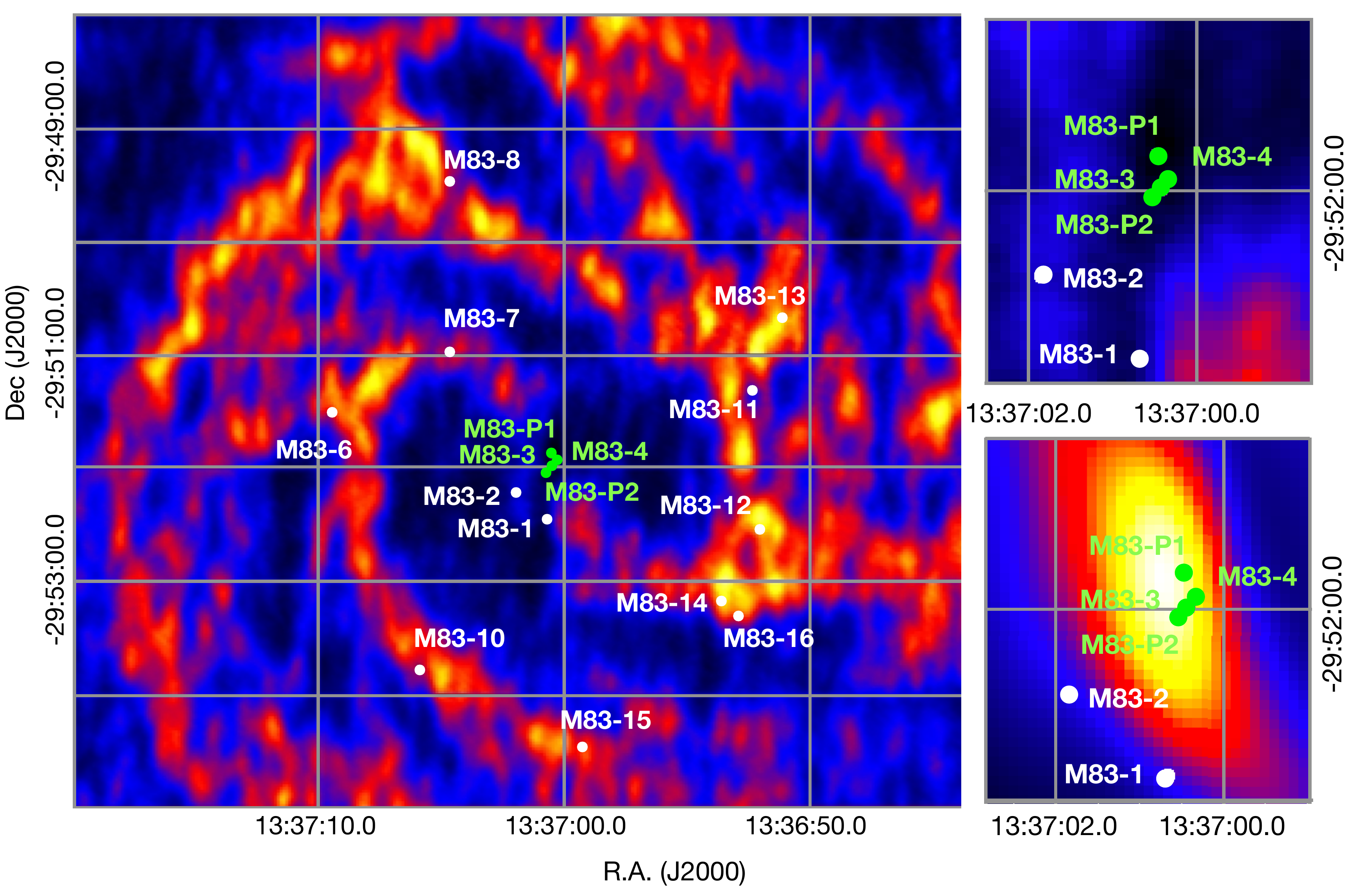}}
      \caption{Left: 21-cm \ion{H}{1} map from THINGS \citep{wal08}. We show the location of the COS M83 pointings with white- and green-font labels. Note that with the exception of four pointings in the nuclear region of this spiral galaxy, most of the targets are located in regions with strong \ion{H}{1} emission. Top right: 21-cm \ion{H}{1} zoom-in of the nuclear region of M83, 55\arcsec $\times$ 50\arcsec. Green dots show the location of the COS pointings with low column densities, $\log$[$N$(\ion{H}{1})]$<$ 20.0 dex. White dots show the location of the pointings with $\log$[$N$(\ion{H}{1})]$>$ 20.0 dex. Bottom right: 55\arcsec $\times$ 50\arcsec CO emission map of the nuclear region of M83 by \citet{hir18} matching the resolution and pixel scale of the \ion{H}{1} map.}
         \label{fig:H_maps}
   \end{figure*}

 To further investigate this anti-correlation between \ion{H}{1} and molecular gas at the center of M83 we inspected the integrated 21-cm \ion{H}{1} map observed and calibrated as part of The \ion{H}{1} Nearby Galaxy Survey\footnote{\href{http://www.mpia.de/THINGS/Overview.html}{http://www.mpia.de/THINGS/Overview.html}} \citep[THINGS,][]{wal08}. In Figure \ref{fig:H_maps} we show a map of the atomic hydrogen from THINGS using the Very Large Array (VLA) and a synthesized beam of 10.4\arcsec $\times$ 5.6\arcsec. The archival image was made with natural weighting of the visibilities. In the top right image of Figure \ref{fig:H_maps} we show a zoomed-in version of the nuclear region of M83. We mark with green circles the location of the four M83 pointings with $\log$[$N$(\ion{H}{1}) cm$^{-2}$]$<$ 20.0. We also show with white dots two pointings with $\log$[$N$(\ion{H}{1}) cm$^{-2}$]$>$ 20.0. From the \ion{H}{1} map it is clear that a depletion of neutral hydrogen is present in the center of this spiral galaxy. \par
 We contrast these results with the CO observations obtained with the Atacama Large Millimeter/submillimeter Array (ALMA) with an extremely high beam resolution of  2.03\arcsec $\times$ 1.15\arcsec and published by \citet{hir18}. We obtained the calibrated CO map resulting after applying a mask to discard velocity pixels dominated by noise (kindly provided by A. Hirota). However, in order to perform a direct comparison with the lower-resolution \ion{H}{1} map, we convolve the high-resolution CO map with a kernel created using the Gaussian profiles from both maps, \ion{H}{1} and CO. As a last step we place the CO image in the same pixel scale as that from the \ion{H}{1} map. The final CO map is shown in the lower right panel in Figure \ref{fig:H_maps}. Similar to the zoomed frame of the \ion{H}{1} map, we also show in this CO image the location of the COS pointings with low column densities of \ion{H}{1} with green dots. We confirm the strong contrast between the depletion of atomic hydrogen gas, and the excess of molecular gas at the core of M83. \par
 Neutral atomic hydrogen has been observed to be depleted within the inner regions of many spiral galaxies \citep{mor78,sag91,cro00,cro07}, however, the precise reason for this depletion is not completely understood. The most natural explanation would be the conversion of atomic gas to molecular gas in regions with high metallicities and dust contents. Since molecular gas is the primary driver of star-formation, star-forming regions are known to have higher molecular gas content \citep{kum20}, naturally explaining the depletion in atomic gas in the center of M83. Moreover, given that H$_{2}$ molecules are created on the surface of dust, one crucial parameter regulating the formation of H$_{2}$ is the amount of dust, which is assumed to be proportional to the metallicity \citep{hon95}. This trend is clear in the work of \citet{cas17}, where they find a high concentration of dust in the core of M83, along with a depletion of \ion{H}{1} in these same regions. Considering that in the nucleus of M83 we find the highest metallicities (see Section \ref{sec:grad}), a scenario where \ion{H}{1} is converted to H$_{2}$ is the most reasonable explanation for the strong depletion observed in our \ion{H}{1} column densities. Lastly, the column densities of $\log$[$N$(\ion{H}{1}) cm$^{-2}$]$\sim$ 21.0 observed just outside of the nuclear region of M83 correspond to the usual threshold where local galaxies begin to experience the \ion{H}{1}--H$_{2}$ transition \citep{sch01,kru09, kru09b}. \par

In the MW similar \ion{H}{1} voids have been observed in the inner Galaxy \citep{loc84,loc16}. It has been proposed that the lack of \ion{H}{1} in the central regions of the MW, compared to the rest of the Galactic disk, might hint at an excavation by Galactic winds \citep{bre80,loc16}. In addition to the scenario provided above, a second explanation for the depletion of \ion{H}{1} observed in the center of M83 could be the excavation by galactic winds, as observed in the MW. This scenario would be supported by studies confirming the existence of an ongoing starburst in the center of this spiral galaxy \citep{dop10, wof11} causing high cluster formation efficiencies (from about $\sim$26\% in the inner region to 8\% outside of this region) and a significant steepening of the initial cluster mass function in the inner regions of M83 \citep{ada15}. The stellar feedback from these young massive star clusters in the core of M83 can generate energetic winds capable of ejecting large fractions of the neutral gas. Furthermore, through the analysis of high-resolution hydrodynamical simulations of the multiphase components of galactic winds, \citet{sch17} find that momentum does not transfer efficiently, and therefore the momentum in the galactic winds is unable to accelerate the dense phase to the wind velocity, failing to entrain the cool dense gas. Until recently, the results from \citet{sch17} were supported by the lack of evidence of cold dense molecular gas in the Galactic nuclear wind. Although this same process could possibly explain the excess in molecular gas observed in the nuclear regions of M83, a recent study by \citet{dit20} reports for the first time on the detection of molecular gas outflowing from the center of the MW. \citet{dit20} confirm that their results pose a challenge for current theoretical models of galactic winds in regular SFGs, as no process is currently able to explain the existence of fast-moving molecular gas in the MW nuclear wind. \par

\subsection{Metallicity gradients in the multi-phase gas and stellar component}\label{sec:grad}
The discovery of the inhomogeneity of metals in the ISM throughout the MW \citep{sha83} has become a fundamental concept in our understanding of galaxy interactions, accretion, mergers and gas flows. Studies of nearby spiral galaxies have shown relatively higher metallicities in the inner regions compared to the outer disk, with gradients typically of the order of $\sim-$0.05 dex kpc$^{-1}$ \citep{pil12}. Most of the studies investigating metallicity gradients in local galaxies have focused on their ionized-gas component \citep[\ion{H}{2} regions, e.g.,][]{wal97,bre05,bre07,kew10,ho15,bre19,kre19}, and to a lesser degree on the stellar component \citep{mol99,car08,san11,dav17}. In contrast, metallicity gradients imprinted in the neutral gas of nearby galaxies have been rather unexplored. \par
\begin{figure*}
   	  \centerline{\includegraphics[scale=0.5]{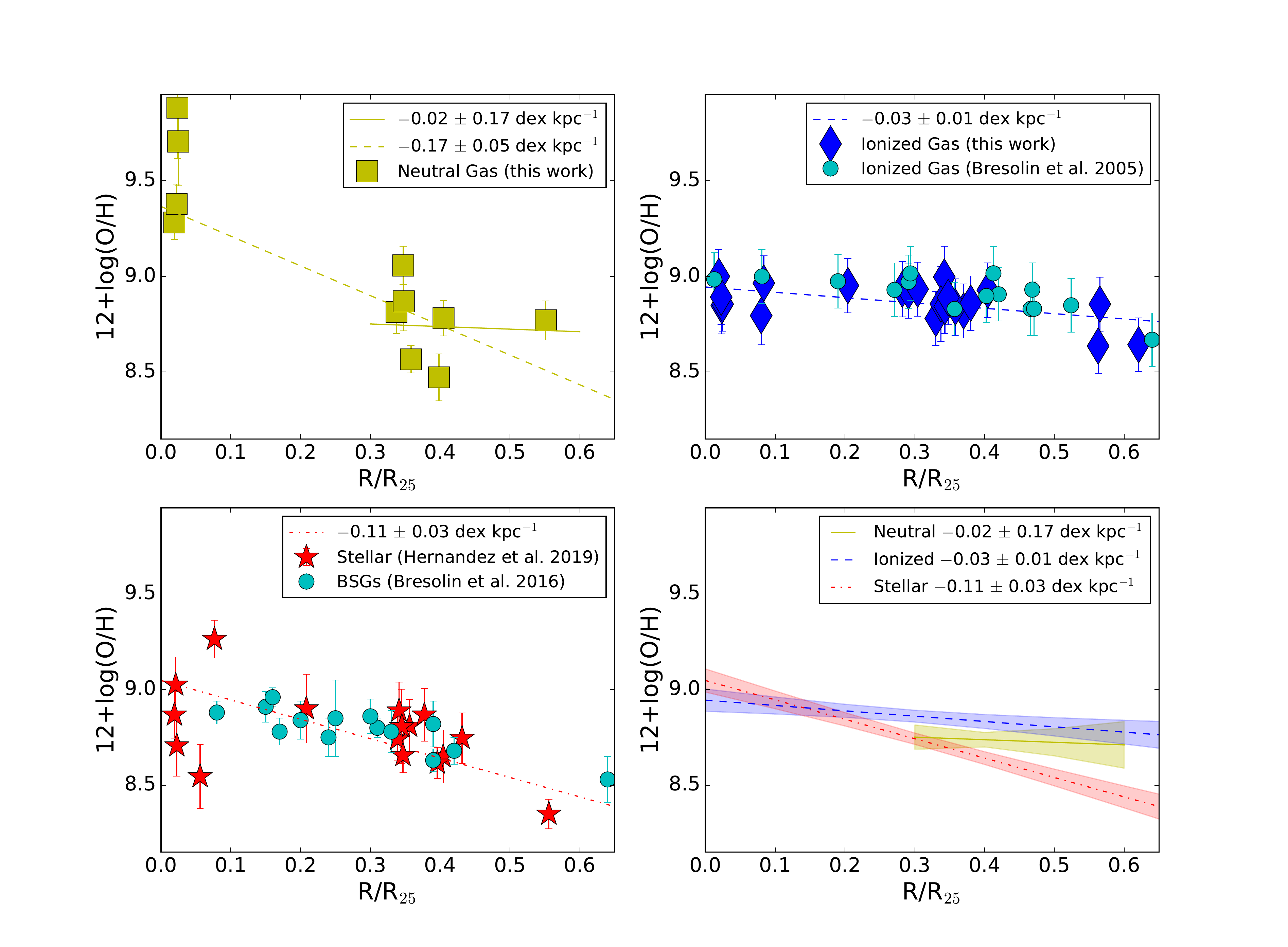}}
      \caption{\textit{Upper left:} Oxygen abundances inferred for the neutral-gas component. The yellow dashed line represents the metallicity gradient inferred for the whole galactocentric range, including the pointings in the nuclear region. The solid yellow line represents the gradient calculated after the exclusion of the pointings in the center of M83. We point out that the lower limits have been excluded both from the figure and when applying the linear regressions. \textit{Upper right:} The ionized-gas abundances calculated as part of this work using the O3N2 calibration by \citet{pet04} are shown in blue diamonds. We show with a dashed blue line the metallicity gradient for the ionized-gas component using our inferred abundances. For comparison, we include the metallicities calculated using the same O3N2 calibration and the \ion{H}{2}-region sample by \citet{bre05}. \textit{Lower left:} Red stars represent the M83 stellar metallicities as measured by \citet{her19}. The dotted dashed red line shows the inferred metallicity gradient. We compare these measurements with those from \citet{bre16} using BSGs, shown as cyan circles. \textit{Lower right:} We show the inferred metallicity gradients, solid yellow line for the neutral gas, dashed blue line for the ionized gas, and dotted dashed line for the stellar populations. The corresponding 1$\sigma$ confidence intervals are shown as shaded regions.}
         \label{fig:Z_gradient}
   \end{figure*}

In this section we will describe in detail the different metallicity trends presented in Figure \ref{fig:Z_gradient} as inferred from the neutral gas (top left), ionized gas (top right), and stellar populations (lower left) in M83. \par


Assuming a single gradient (one without breaks), we also show in Figure \ref{fig:Z_gradient} linear regressions for the different metallicity components. As previously discussed, given the high metallicities in the core of the galaxy, particularly in the neutral gas, we find a steep gradient of the order of $-$1.6 $\pm$ 0.4 dex $R_{\rm 25}^{-1}$ (dashed yellow line in upper left panel). We estimate a slightly shallower gradient for the stellar component of $-$1.0 $\pm$ 0.3 dex $R_{\rm 25}^{-1}$ (dotted-dashed red line in lower left panel), and a much shallower one for the ionized-gas component of $-$0.2 $\pm$ 0.1 dex $R_{\rm 25}^{-1}$ (dashed blue line in upper right panel). We note that to obtain an accurate view of the metallicity gradients, we exclude the lower limit values when fitting the neutral-gas measurements.  \par

\subsubsection{Neutral-gas metallicity gradient}
\begin{figure}
   	  \centerline{\includegraphics[scale=0.4]{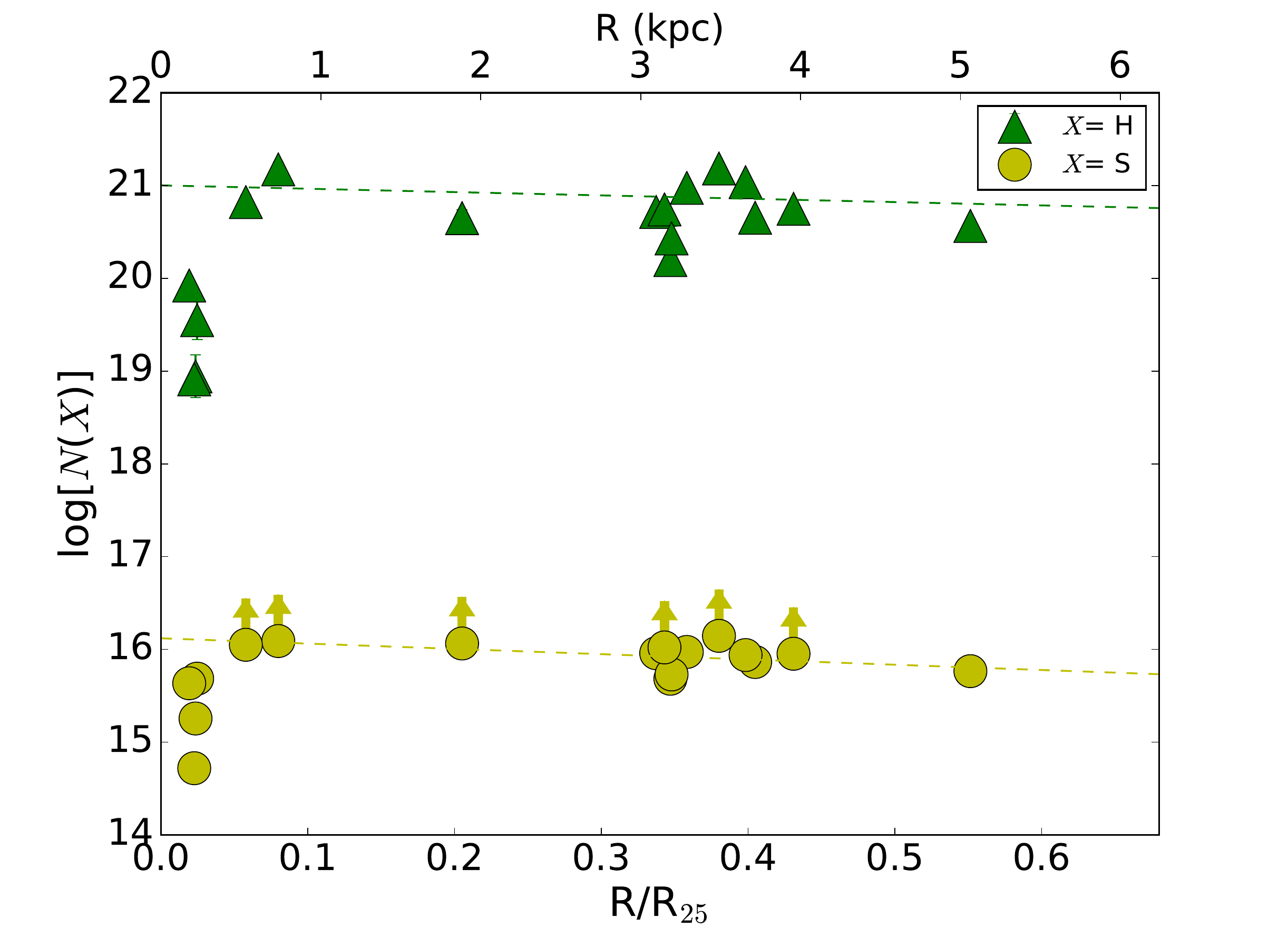}}
      \caption{Column densities for hydrogen (green) and sulfur (yellow) as a function of galactocentric distance. To help guide the eye we show linear regressions for those points at R/R$_{\rm 25}> 0.02$ excluding the sightlines with depleted column densities in the nuclear region, and including the lower limits of \ion{S}{2}. }
         \label{fig:SvsHcol}
   \end{figure}

The steep gradient observed in the neutral gas  (upper left panel in Figure \ref{fig:Z_gradient}, $-$1.6 $\pm$ 0.4 dex $R_{\rm 25}^{-1}$  or $-$0.17 $\pm$ 0.05 dex kpc$^{-1}$) appears to be primarily driven by the low column densities of \ion{H}{1} in the nucleus of the galaxy (discussed in Section \ref{sec:HI}). In an effort to better understand the derived neutral-gas metallicities in the core of M83 we display the column densities of H and S as a function of galactocentric distance in Figure \ref{fig:SvsHcol}. Figure \ref{fig:SvsHcol} shows that the neutral metals, traced by S, are depleted at the center, to a lesser degree than H. To help guide the eye we apply linear regressions to the column densities of H and S at R/R$_{\rm 25}> 0.02$ excluding the highly depleted pointings, and including the lower limits in S to have a better galactocentric coverage. We note that the inclusion of the lower limits in S returns a gradient which should be considered as a lower limit as well. \par 
Following the trends in Figure \ref{fig:SvsHcol}, shown as dashed lines, and assuming the radial profile of H should hold in the center of M83, we would expect column densities of H of the order of $\log$[$N$(H) cm$^{-2}$] $\sim$ 21.0, and instead we measure $\log$[$N$(H) cm$^{-2}$] $\sim$ 18.9. This implies that $\sim99\%$ of the neutral H has been depleted. Using S as a metallicity tracer in the neutral gas, from Figure \ref{fig:SvsHcol} we would expect column densities of $\log$[$N$(S) cm$^{-2}$] $\gtrsim$ 16.2, and instead measure column densities as low as $\log$[$N$(S) cm$^{-2}$] $\sim$ 14.7. Note that for S, we basically assume that the S/H ratio is approximately constant. The observed trend hints at a depletion in the neutral metals of $\gtrsim97\%$. One possible explanation for the differences observed in the fraction of depleted neutral H and S might be linked to the high concentration of molecular gas in the nucleus of M83 discussed in Section \ref{sec:HI}.\par
Given that the medium in the center of this spiral galaxy is mainly in molecular form \citep{cas17}, it is possible that a significant fraction of the measured S might be tracing the CO-dark H$_{2}$ gas, providing a biased estimate of the metals in the neutral gas. In the neutral gas, ions such as C$^+$, Si$^+$ and O$^0$ can exist in both the \ion{H}{1} and H$_{2}$ phases, particularly in regions where CO is photodissociated and molecular hydrogen is self-shielded (and shielded by dust) from UV photodissociation, typically called CO-dark H$_{2}$ gas \citep{mad97, wol10}. Since the ionization potential of S$^+$ is comparable to that of C$^+$, Si$^+$ and O$^0$, we might also expect S$^+$ to coexist in the CO-dark H$_{2}$ phase. In such a scenario, and based on the calculations presented above, we estimate that from the measured column density of $\log$[$N$(S) cm$^{-2}$] $\sim$ 14.7 and assuming a similar fraction of S and H, i.e., 99$\%$ depletion, we can then infer that the column density of S arising from the CO-dark H$_{2}$ phase is of the order of $\log$[$N$(S) cm$^{-2}$] $\sim$ 14.6. If this hypothesis proves to be accurate, we are possibly probing an intercloud or clumpy medium, where the matter is in the molecular (CO-dark) but diffuse phase. If we assume an efficient conversion from \ion{H}{1} to H$_{2}$, we can expect a column density for the molecular hydrogen of the order of $\log$[$N$(H$_{2}$) cm$^{-2}$]$\sim$ 21. We contrast this expected column density of H$_{2}$, with those observed from high-resolution CO maps (with a synthesized beam of 2.0$\arcsec$ $\times$ 1.1$\arcsec$ in full width at half maximum) by \citet{fum18} where at the peak of the emission they observe column densities of the order of $\log$[$N$(H$_{2}$) cm$^{-2}$] $\gtrsim$ 22, and between the peaks values of $\log$[$N$(H$_{2}$) cm$^{-2}$]$\sim$ 20-21. We note that these column densities of H$_{2}$ are obtained assuming the Galactic conversion factor ($X_{\rm CO}$) by \citet{dam01}. These values would suggest that the medium in the center of M83 could be $\sim$10 $\%$ CO-dark H$_{2}$ gas towards the peaks observed in the CO maps, and 100$\%$ CO-dark H$_{2}$ gas between the peaks. This hypothesis could be put to test by probing H$_{2}$ in absorption, which should trace both phases, the CO-dark and CO-bright. Finally, if such a high fraction of our inferred S column densities originates from the CO-dark H$_{2}$ gas, we may assume that in active and star-forming environments dominated by molecular gas, the typical neutral-gas metallicity tracers (e.g., S and O) provide a biased view of the total metal contents in the neutral ISM. \par

Regarding the neutral-gas trend, to confirm if indeed there is a gradient present outside the nuclear region of M83, we analyze the gradient of the neutral gas excluding those pointings in the nuclear region. We find instead a relatively flatter neutral-gas metallicity gradient of $-$0.02 $\pm$ 0.17 dex kpc$^{-1}$ (shown as solid yellow line in the upper right panel in Figure \ref{fig:Z_gradient}). Although we provide a value for a possible gradient in the neutral gas, outside of the nucleus of M83, we note that our measurement is limited by a low number of pointings available, primarily clustered around R $\sim$ 0.4 R$_{25}$. A larger sample, spanning a much broader galactocentric range, will be needed to reduce the uncertainties in our measurements allowing us to draw firmer conclusions and better characterize the metallicity gradient of the neutral gas. \par
Overall, our work shows that caution must be taken when studying the abundance gradients of spiral galaxies, as the intense activity in the center of these star-forming galaxies (outflows, high star-formation rates, high concentrations of molecular gas) can strongly bias the inferred neutral-gas gradients.  

\subsubsection{Ionized-gas metallicity gradient}
In the upper right panel in Figure \ref{fig:Z_gradient} we show with blue diamonds our inferred ionized-gas metallicities and with a blue dashed line the corresponding metallicity gradient,  $-$0.3 $\pm$ 0.1 dex $R_{\rm 25}^{-1}$ or $-$0.03 $\pm$ 0.01 dex kpc$^{-1}$. We find an excellent agreement between our inferred abundances and  those calculated by \citet{bre16} using the O3N2 calibration by \citet{pet04} and the line fluxes from \citet{bre05}, shown as cyan circles. \citet{bre16} calculate an ionized-gas metallicity gradient of $-$0.24 $\pm$ 0.06 dex $R_{\rm 25}^{-1}$, similar to that measured here. \par
It is known that ionized-gas metallicity gradients in local galaxies are correlated with Hubble type, particularly bar strength and merging episodes. Studies have shown that relatively flatter gradients are found for barred galaxies and merging pairs than other types \citep{kew10,rup10}. In a recent study by \citet{ho15}, they inferred metallicity gradients from \ion{H}{2} regions in 49 local star-forming galaxies and find a strong correlation between their stellar mass and the observed gradients. They provide a local benchmark of metallicity gradients where galaxies with stellar masses of log($M_{*}/M_{\odot}$) $>$ 9.6 are expected to have gradients of the order of $-$0.026 $\pm$ 0.002 dex kpc$^{-1}$, in agreement with the median slope inferred for an isolated spiral control sample by \citet{rup10b}. M83 has a stellar mass of log($M_{*}/M_{\odot}$) $=$ 10.55 \citep{bre16} and, based on our analysis, an ionized-gas metallicity gradient of $-$0.03 $\pm$ 0.01 dex kpc$^{-1}$, well within the range of the masses defined by \citet{ho15} and with a gradient comparable to their benchmark value. The common slopes between the sample in \citet{ho15} and M83 imply that, in general, disk galaxies evolve in a similar manner when developing their disks, possibly following an inside-out disk growth model \citep{san14}. 	\par
 \par

\subsubsection{Stellar metallicity gradient}
\citet{her19} estimate a stellar metallicity gradient for M83 of $-$0.38 $\pm$ 0.20 dex $R_{\rm 25}^{-1}$, the main difference between their study and the work presented here is the galactocentric ranges considered. \citet{her19} examine the stellar metallicities at $R/R_{\rm 25} < $ 0.5, excluding the low metallicity cluster at $R/R_{\rm 25} \sim $ 0.55, whereas in this work we include all of the pointings when calculating the stellar gradient. The main driver for excluding the stellar cluster at $R/R_{\rm 25}$$\sim$ 0.55 is justified by a proposed break in the abundance gradient \citep{her19}. Given that such a break is less visible in the nebular metallicities calculated as part of this work, we decide to include this stellar cluster at $R/R_{\rm 25}$$\sim$ 0.55 in our analysis. 
\par

In the lower left panel of Figure \ref{fig:Z_gradient} we compare the stellar metallicities from our YMC sample (red stars) with those from the BSG sample (cyan circles) by \citet{bre16}. To accurately compare the results from the two studies, we homogenize the inferred metallicities to a single abundance scale using equation 6 in \citet{her19}. We highlight the clear agreement between the two independent studies. \par
In the stellar metallicity study by \citet{bre16}, they compare their measurements with those from the ionized gas using different strong-line indicators. In general, \citet{bre16} find that the \ion{H}{2} abundance gradients are significantly shallower than those from the blue supergiants or those obtained from the direct T$_{e}$-method. In spite of the shallower gradient obtained using the O3N2 calibration, they suggest that the radial metallicity distribution and the scatter in their stellar abundances resemble those observed in the \ion{H}{2} abundances inferred with O3N2 diagnostic. This pattern is similar to what we see in the work presented here, where the stellar metallicity gradient is slightly steeper (lower left panel in Figure \ref{fig:Z_gradient}) than that inferred for the ionized gas (upper right panel in Figure  \ref{fig:Z_gradient}) using the O3N2 calibration.\par
In addition to observing a similar trend to that in \citet{bre16} when comparing our ionized-gas gradient with the stellar metallicity gradient, we also highlight the agreement between our inferred values. \citet{bre16} measure an abundance gradient of $-$0.66 $\pm$ 0.13 dex $R_{25}^{-1}$ using BSGs, and $-$0.81 $\pm$ 0.57 dex $R_{25}^{-1}$ from \ion{H}{2} regions using the T$_{e}$-based method, well within the uncertainties of our stellar metallicity gradient, $-$1.0 $\pm$ 0.3 dex $R_{\rm 25}^{-1}$. \par


\subsubsection{Comparison of Global Metallicity Gradients}
Lastly, in the lower right panel in Figure \ref{fig:Z_gradient} we compare the metallicity gradients from the three different components: neutral-gas (solid yellow line), ionized-gas (dashed blue line), and stellar (dotted dashed red line).  Overall, our work proves to be consistent with previous studies where metallicity gradients inferred from stellar populations are steeper than those measured in the ionized gas using strong-line calibrators. Furthermore, comparing the observed gradient for the stellar populations ($-$0.11 $\pm$ 0.03 dex kpc$^{-1}$) in M83 to the benchmark gradient for nearby SFGs by \citet{ho15}, we find that the stellar value is much higher than what is expected from the ionized-gas component, $-$0.026 $\pm$ 0.002 dex kpc$^{-1}$.  On the other hand, the benchmark gradient and that inferred from the neutral-gas component, $-$0.02 $\pm$ 0.17 dex kcp$^{-1}$, agree within the large uncertainties of the inferred neutral-gas metallicity gradient. \par

Our work shows that it is critical to examine in detail the effects and activity in the nuclear regions of star-forming galaxies when studying metallicity gradients imprinted in the neutral-gas component, as these might be strongly biased if the dominant phase is molecular. And finally, we highlight that even when the inferred gradients for the different components appear to be dissimilar, we note that outside of the central regions the abundances of the multiphase gas and stellar populations appear to be more homogenized, with similar metallicities (within the uncertainties of our measurements) throughout the different components, e.g., at R $>$ 0.2 R$_{25}$. We discuss the metallicities of the individual pointings and their different components in the following section. \par



\subsection{Co-spatial comparisons: Stellar, Neutral- and Ionized-gas Metallicities}
We now compare the metallicities of the multi-phase ISM with those from the stellar populations for each individual COS pointing. This provides a co-spatial comparison on the small galactic scales of the star clusters ($\sim$100 pc). \par

\begin{figure*}
   	  \centerline{\includegraphics[scale=0.32]{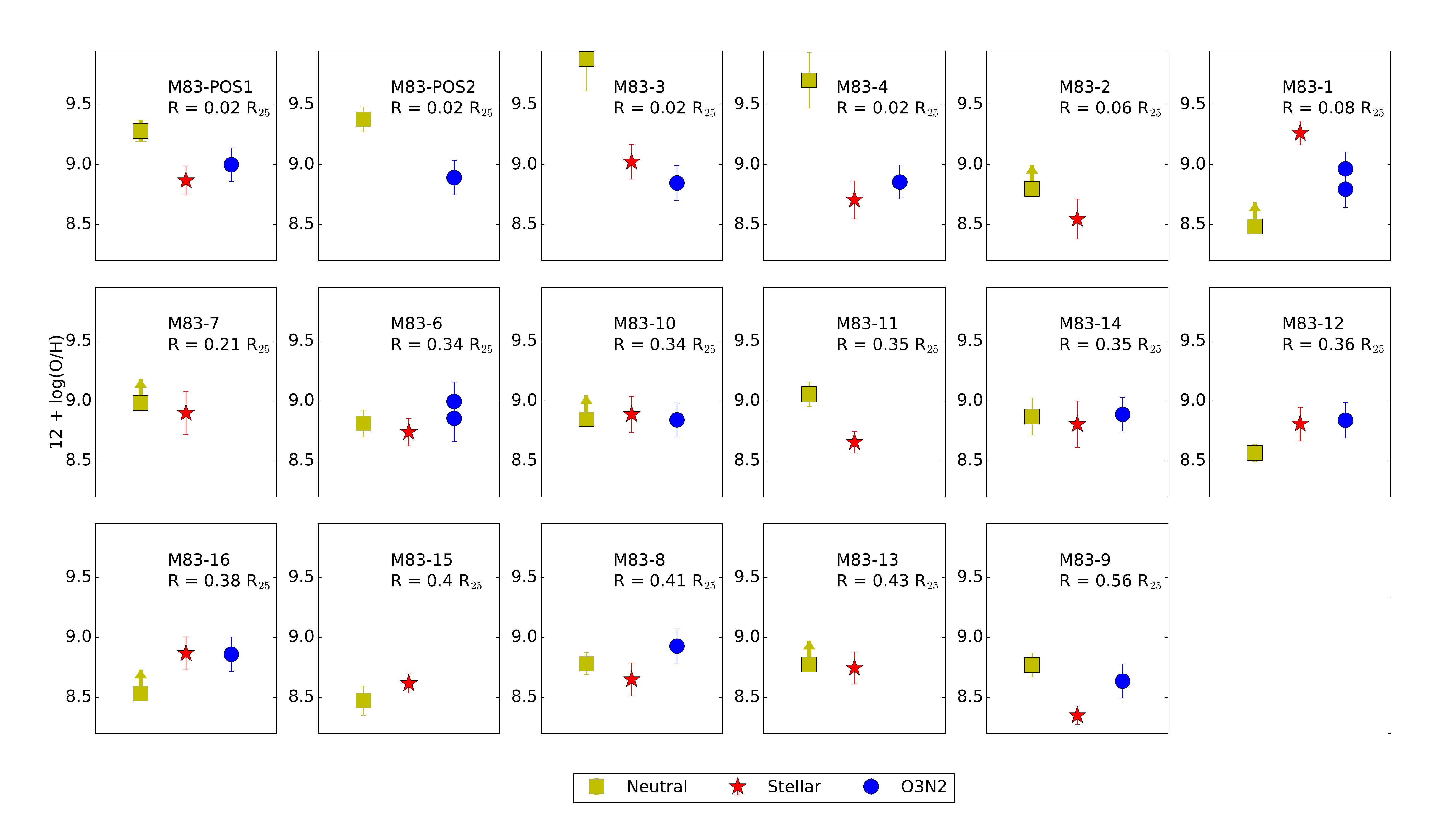}}
      \caption{Metallicities for the individual M83 pointings. Yellow square show the oxygen abundances of the neutral gas inferred from the COS observations. We indicate with upper arrows the lower limit values. Red stars display the stellar abundances as measured by \citet{her19}. Blue circles show the ionized-gas oxygen abundances measured from the LBT and MUSE data using the O3N2 \citep{pet04} calibration.}
         \label{fig:Z_ind}
   \end{figure*}

In Figure \ref{fig:Z_ind}, we display individual panels comparing the metallicities of each phase, neutral-gas (yellow squares), stellar (red stars), and ionized-gas (blue diamonds). We can see that with the exception of two pointings, {M83-1} and {M83-9}, the stellar metallicities and the nebular metallicities agree within their uncertainties. We note the contrast between these two pointings, M83-1 is close to the nuclear region of M83, and M83-9 is the pointing at the largest galactocentric distances in our sample. In general, the clear agreement we observed in our study supports a scenario where the gas surrounding these young populations of stars (average age $\sim$7 Myr), particularly the hot gas, is not instantaneously enriched by the most massive stars. We arrive at a similar conclusion to that in \citet{chi19}, where the agreement between the abundances of the stellar populations and the ionized gas in a SFG sample at low ($<$0.2) and high ($\sim$2) redshifts indicate that the gas adjacent to the young stars is enriched at longer timescales than the lifetimes of the most massive stars. Our work shows that at least on small galactic scales of $\sim$100 pc it takes $>$10$^{7}$ yr to fully mix the newly processed metals from the massive stars into the ISM. 
These longer than expected timescales imply caution must be taken when assuming instantaneous recycling approximations when generating galactic chemical evolution models \citep{kob97}. \par

We find a similar trend for the neutral gas. The metallicities inferred for the neutral-gas component of M83 also show an overall agreement with those from both the stellar, and the ionized-gas component, once we accurately consider the lower limits. The exception to this general trend is found primarily in the nucleus of M83. We find that particularly at the core of M83, the metallicity of the neutral gas is more enhanced than that for the stellar and ionized-gas components. This is clearly seen in the first four top panels in Figure \ref{fig:Z_ind}, M83-POS1, M83-POS2, M83-3 and M83-4. This enhancement in the metallicity of the neutral gas is discussed above in Section \ref{sec:grad}. Given that in the center of M83 the dominant phase is clearly molecular (Figure \ref{fig:H_maps}), the S column densities observed in these four pointings trace both the neutral and (CO-dark) molecular gas, providing a biased view of the neutral-gas metallicities.\par

Lastly, to further investigate a possible spatial difference between the neutral gas and the stellar populations, we look at the radial velocities of the neutral-gas and stellar components listed in Table \ref{tab:vel}. We find that the average radial velocity difference between the neutral gas and the stars is $\Delta \bar v$ = $v_{\rm neutral} - v_{\rm stellar}$ = $-$13.0 km s$^{-1}$ with a standard deviation of $\sigma$ = 34.4 km s$^{-1}$. For pointings M83-3 and M83-4 we find a radial velocity difference of $\Delta v$ = $-$42.9 km s$^{-1}$ and $-$40.4 km s$^{-1}$, respectively, within the standard deviation of the mean $\Delta \bar v$ in our sample. For M83-POS1, we estimate a larger velocity difference between the two components of $\Delta v$ = $-$65.1 km s$^{-1}$, potentially indicating that the neutral gas along the line of sight of M83-POS1 is occupying a spatially different region than that of the star cluster. \par
Overall, our co-spatial comparison shows that the three different components, neutral gas, ionized gas, and stellar populations, are chemically homogeneous on small scales ($\sim$100 pc), with the exception of the nuclear region in M83. In the galactic center we are unable to make a direct comparison between the metallicity of the neutral gas and that of the ionized gas and stellar populations, as the inferred neutral-gas abundances are strongly biased due to the dominant molecular phase in these regions.
  
\begin{table}
\caption{Radial velocities for the neutral-gas and stellar component.}
\label{tab:vel}
\centering 
\begin{tabular}{ccc}
\hline \hline
Target & $v_{\rm neutral}$ & $v_{\rm stellar}^{a}$\\
  & (km s$^{-1}$) &  (km s$^{-1}$) \\
\hline\\
        M83-1& 497 $\pm$ 13	& 468 $\pm$ 43\\	 	 
	M83-2&517 $\pm$ 9 & 503 $\pm$ 8\\ 	 	 	  	 
	M83-3&465 $\pm$ 13 & 508 $\pm$ 9\\	 	 	  	  	 	 
	M83-4&490 $\pm$ 15	&530 $\pm$ 3\\ 	
	M83-5& -- & -- \\
	M83-6&447 $\pm$ 9&	443 $\pm$ 7\\ 	  	 
	M83-7& 477 $\pm$ 7&	455 $\pm$ 17\\
	M83-8&381 $\pm$ 15	&464 $\pm$ 12\\	 	 	 		 	  
	M83-9& 442 $\pm$ 12	&407 $\pm$ 22\\ 
	M83-10& 504 $\pm$ 14 & 514 $\pm$ 1\\ 
	M83-11& 513 $\pm$ 8	& 524 $\pm$ 22\\ 
	M83-12& 562 $\pm$ 7	 & 539 $\pm$ 13\\ 
	M83-13& 523 $\pm$ 9	& 517 $\pm$ 1\\ 
	M83-14& 512 $\pm$ 8	 & 560 $\pm$ 15\\ 
	M83-15& 538 $\pm$ 20 & 570 $\pm$ 4\\ 
	M83-16& 530 $\pm$ 6	& 539 $\pm$ 4\\ 		 	 	 	 	 
	M83-POS1& 430 $\pm$ 20	& 495 $\pm$ 10\\	 	 	 	 	 
	M83-POS2 & 493 $\pm$ 19 & -- \\	 	 	 	 	 
\hline
\end{tabular}
\begin{minipage}{15cm}~\\
\textsuperscript{$a$}{Taken from \citet{her19}.}\\
 \end{minipage}
\end{table}

\section{Conclusion}\label{sec:con}
The work presented here aims at providing a comparative study of the multi-phase gas and the stellar component in the nearby spiral galaxy M83. We analyze HST/COS, LBT/MODS, and VLT/MUSE observations to estimate the metallicities of the neutral gas, ionized gas and stellar populations throughout the disk of this metal-rich galaxy. We summarize our results as follows:
   \begin{enumerate}
   \item We find a clear depletion of \ion{H}{1} gas as observed from the \ion{H}{1} column densities in the core of M83. At galactocentric distances of R/R$_{25}$$<$ 0.02 (R $<$ 0.18 kpc) we estimate column densities of the order of log[$N$(\ion{H}{1}) cm$^{-2}$] $<$ 20.0. At galactocentric distances of R/R$_{25}$$>$ 0.02 we find that the \ion{H}{1} column densities increase to values of log[$N$(\ion{H}{1}) cm$^{-2}$] $\sim$ 21.0, typical of spiral galaxies. 
   \item After comparing the \ion{H}{1} and molecular gas maps, we find a clear anti-correlation at the core of M83 where the region with depleted \ion{H}{1} column densities shows a significant excess of molecular gas. 
   \item Using the O3N2 calibration by \citet{pet04} we measure a metallicity gradient of $-$0.03 $\pm$ 0.01 dex kpc$^{-1}$ for the ionized gas, comparable to the local benchmark of metallicity gradients of nearby star-forming galaxies by \citet{ho15}, implying that disk galaxies evolve in a similar manner, following an inside-out model.
   \item Our work shows that outside of the nuclear region of M83 the metallicity of the neutral-gas, ionized-gas and stellar populations are comparable (within the uncertainties of our measurements) and more homogenized than in the nucleus of the galaxy. These findings call for caution when studying abundance gradients, particularly for the neutral-gas component, which can be strongly biased by the observed metallicities in the center as these are greatly affected by the efficient conversion from atomic to molecular gas. 
   \item We find a slightly steeper stellar metallicity gradient, $-$0.11 $\pm$ 0.03 dex kpc$^{-1}$, compared to that observed in the ionized gas, $-$0.03 $\pm$ 0.01 dex kpc$^{-1}$, using the strong-line calibration of O3N2. This trend is similar to that observed in previous studies, where the abundance gradients inferred from strong-line calibrations are shallower than those observed from stellar populations. 
   \item A co-spatial comparison of the metallicities of the multi-phase gas and the stellar populations show excellent agreement outside of the nucleus of the galaxy. This hints at a scenario where on small galactic scales ($\sim$100 pc) it takes longer than the lifetime of the most massive stars ($\sim$10 Myr) to fully mix the newly synthesized metals. 
   \end{enumerate}
  Overall, we observe homogeneous metallicities in the multi-phase gas and stellar component on small scales similar to those of stellar clusters in the high-metallicity environment of M83. Studies similar to that detailed in this paper are critical for validating this trend in a variety of other environments in the Local Universe. Lastly, we highlight that studies like the one presented here would benefit greatly from multiphase high spatial resolution simulations to further understand the feedback mechanisms and processes taking place as well as mixing timescales in much more detail. 

\acknowledgments

These data are associated with the HST GO programs 14681, 11579, and 15193 (PI: A. Aloisi). Support for this program was provided by NASA through grants from the Space Telescope Science Institute. Some of the data presented in this paper were obtained from the Mikulski Archive at the Space Telescope Science Institute (MAST). This paper uses data taken with the MODS spectrographs built with funding from NSF grant AST-9987045 
and the NSF Telescope System Instrumentation Program (TSIP), with additional funds from the Ohio Board of 
Regents and the Ohio State University Office of Research. 
This paper made use of the modsIDL spectral data reduction pipeline developed by Kevin V Croxall in part 
with funds provided by NSF Grant AST-1108693. 
This work was based in part on observations made with the Large Binocular Telescope (LBT). 
The LBT is an international collaboration among institutions in the United States, Italy, and Germany. 
The LBT Corporation partners are: the University of Arizona on behalf of the Arizona university system; 
the Istituto Nazionale di Astrofisica, Italy; the LBT Beteiligungsgesellschaft, Germany, representing 
the Max Planck Society, the Astrophysical Institute Potsdam, and Heidelberg University; the Ohio State 
University; and the Research Corporation, on behalf of the University of Notre Dame, the University of 
Minnesota, and the University of Virginia. CAFG was additionally supported by NSF through grants AST-1715216 and CAREER award AST-1652522; by NASA through grant 17-ATP17-0067; and by a Cottrell Scholar Award from the Research Corporation for Science Advancement. This work made use of THINGS, `The \ion{H}{1} Nearby Galaxy Survey'.

%

\vspace{5mm}
\facilities{HST(COS)}

%
\software{CLOUDY \citep{fer17}, VoigtFit v.0.10.3.3 \citep{kro18}, CALCOS pipeline (v.3.3.4)}



\appendix
\section{Observed LBT/MODS targets}
We present the coordinates of the MODS pointings observed as part of PID: LBT-2018A- I0037-0 (PI: Skillman and Berg), and analyzed as part of our work. In Table \ref{tab:mods} we list the coordinates of the MODS slits. 

\begin{table}
\caption{LBT/MODS slit coordinates}
\label{tab:mods}
\centering 
\begin{tabular}{lcc}
\hline \hline
Target & R.A. & Dec \\
& ($^{\circ}$) & ($^{\circ}$) \\
\hline
R1 (M83-9) & 204.2917833 &$-$29.8183306 \\
R2 (M83-8) & 204.2702875 &$-$29.8247167 \\ 
R3 & 204.3253625 & $-$29.8080694 \\
R4 & 204.2265583 & $-$29.8412944\\ 
R5 & 204.2821417 & $-$29.8543528 \\ 
R6 (M83-6) & 204.2900583 & $-$29.8586167 \\ 
R7 & 204.2508208 & $-$29.8640583 \\ 
R8 (M83-1)& 204.2527333 & $-$29.8754000 \\ 
R9 & 204.2849000 & $-$29.8696306 \\ 
R10 & 204.3261792 & $-$29.8812833 \\   
R11 & 204.2286583 & $-$29.8860306 \\ 
R12 & 204.2693667 & $-$29.8498333 \\  
R13 & 204.2330833 & $-$29.8316361 \\ 
R14 & 204.2240375 & $-$29.8126472\\ 
R15 & 204.2445000 & $-$29.7994694 \\ 
\hline
\end{tabular}
\end{table}

\section{Doppler $b$ parameter} \label{app:bval}
As part of the neutral-gas analysis done on the COS observations, the \texttt{VoigtFit} software estimates the Doppler parameter, $b$. In Table \ref{tab:bval} we list the Doppler $b$ values obtained for the different M83 pointings. We do not include the Doppler parameter, $b$,  for the \ion{H}{1} fits, as the line is in the damped part of the COG \citep{jam14}.

\begin{table*}
\caption{$b$ parameters for the M83 pointings.}
\label{tab:bval}
\centering 
\begin{tabular}{ccc}
\hline \hline
Target & \ion{S}{2} & \ion{S}{2}$_{2}$$^{a}$ \\
  & (km s$^{-1}$)  & (km s$^{-1}$)  \\
\hline\\
M83-1&	105.73 $\pm$ 21.55 &--\\	
M83-2&	68.27 $\pm$   9.85	&101.30 $\pm$  28.27\\
M83-3&	139.74 $\pm$  41.58 & --\\	
M83-4&	124.33 $\pm$  20.23&-- \\	
M83-5&	-- & -- \\
M83-6&	123.84 $\pm$  31.97 &-- \\	
M83-7&	79.81 $\pm$  16.72&	53.94 $\pm$  74.77\\
M83-8&	109.62 $\pm$ 19.52&-- \\	
M83-9&	163.28 $\pm$ 17.07 &--	\\
M83-10&	118.69 $\pm$  11.92	&29.30 $\pm$ 23.59\\
M83-11&	96.67 $\pm$   7.49	&-- \\
M83-12&	116.61 $\pm$   8.88 & --\\	
M83-13&	91.19 $\pm$ 10.63	&-- \\
M83-14&	85.17 $\pm$  11.70	 & 54.29 $\pm$  17.56\\
M83-15&	111.33 $\pm$  26.15&	14.33 $\pm$  38.41\\
M83-16&	105.28 $\pm$  11.02	& --\\
M83-POS1&	127.29 $\pm$  15.56&	 88.73 $\pm$  39.19\\
M83-POS2&	127.21$\pm$ 28.32	& --\\
\hline
\end{tabular}
\begin{minipage}{15cm}~\\
 \textsuperscript{$a$}{Multi-component cases. A second \ion{S}{2} component was identified for these pointings.}\\
 \end{minipage}
\end{table*}

\section{Optical emission line fluxes}\label{app:optical}
We present tables detailing the individual emission line measurements for both the LBT/MODS (Tables \ref{tab:flux1}--\ref{tab:flux2}) and VLT/MUSE observations (Tables \ref{tab:flux3}--\ref{tab:flux4}).

	\movetabledown=2.2in	
	\begin{rotatetable}
         \begin{deluxetable}{ccccccccccc}	
	\tablecaption{Emission line measurements (relative to H$\beta$ = 100) for the LBT spectra. Line fluxes (F$_{\lambda}$) are extinction corrected using E(B-V) to calculate I$_{\lambda}$.\label{tab:flux1}}	
\tablehead{ \colhead{Target} & \multicolumn{2}{c}{ \centering $[OII]$ } & \multicolumn{2}{c}{$H\beta$} & \multicolumn{2}{c}{$[OIII]$} &\multicolumn{2}{c}{$[NII]$} & \multicolumn{2}{c}{$H\alpha$} \\ 
\colhead{} & \multicolumn{2}{c}{(3726.03 \r{A})} & \multicolumn{2}{c}{(4861.33 \r{A})} & \multicolumn{2}{c}{(5006.84 \r{A})} & \multicolumn{2}{c}{(6548.03 \r{A})} & \multicolumn{2}{c}{(6562.80 \r{A})} \\
\colhead{} & \colhead{F$_{\lambda}$} & \colhead{I$_{\lambda}$}& \colhead{F$_{\lambda}$} & \colhead{I$_{\lambda}$}& \colhead{F$_{\lambda}$} & \colhead{I$_{\lambda}$}& \colhead{F$_{\lambda}$} & \colhead{I$_{\lambda}$}& \colhead{F$_{\lambda}$} & \colhead{I$_{\lambda}$}}
\startdata
			R1 & $139.25 \pm 35.64 $ &$188.66 \pm 54.43 $ &  $100.00 \pm 3.45 $ & $100.00 \pm 8.06 $ & $67.73 \pm 3.05$ & $64.56\pm10.05$ & $43.39 \pm 4.19 $& $30.73 \pm 3.49 $& $404.77 \pm 15.61$& $286.00 \pm 28.42$ \\
			R2 & $86.23 \pm 4.70 $ & $130.84 \pm 25.07 $ & $100.00 \pm 2.68 $ & $100.00 \pm 12.97 $& $ 9.23 \pm 0.67$& $8.64 \pm 1.38 $& $54.26 \pm 6.02$& $33.79 \pm 6.38$& $460.77 \pm 16.04$ & $286.00 \pm 45.77$\\
			R3 & -- & -- & -- &-- & -- &-- & -- & -- & -- \\
			R4 & $73.17 \pm 1.72$ & $113.85 \pm 12.22$& $100.00 \pm 0.69$& $100.00 \pm 6.37$& $27.55 \pm 0.52$& $25.70 \pm 2.47$& $59.35 \pm 5.58$& $35.92 \pm 4.65$&$ 474.20 \pm 10.27$& $286.00 \pm 25.14$\\
			R5 & $66.09 \pm 5.48$& $100.46 \pm 19.04$ & $100.00 \pm 3.62$ & $100.00 \pm 13.51$& $8.95 \pm 0.84$& $8.38 \pm 1.87$& $54.61 \pm 6.71$& $33.95 \pm 6.80$ &$ 461.68 \pm 20.13$ & $286.00 \pm 49.63$ \\
			R6 &$ 100.67 \pm 9.06$ & $120.61 \pm 29.51$& $100.00 \pm 3.85$& $100.00 \pm  13.22$ & $5.22 \pm 2.84$& $5.08 \pm 2.83$& $40.94 \pm 4.43$ & $33.34 \pm 6.09$ & $351.64 \pm 15.37$ & $286.00 \pm 44.19$ \\
			R7 & $30.31 \pm 6.85$ & $71.36 \pm 15.66$ & $100.00 \pm 0.94$&$100.00 \pm 4.00$& $10.19 \pm 0.39$& $8.90 \pm 0.57$& $4121.12 \pm 4.80$& $45.81 \pm 2.85$ &$761.52 \pm 10.60$& $286.00 \pm 13.49$ \\
			R8 & $41.20 \pm 7.59$ & $79.13 \pm 19.14$& $100.00 \pm 2.21$& $100.00 \pm 12.24$& $7.61 \pm 1.04$& $6.86 \pm 1.24$& $78.39 \pm 7.08$& $37.36 \pm 6.15$& $603.37 \pm 17.91$& $286.00 \pm 39.63$\\
			R9 & $63.45 \pm 5.39$& $137.20 \pm 22.16$& $100.00 \pm 2.43$& $100.00 \pm 8.51$ &$11.64 \pm 0.85$&$10.31 \pm 1.38$& $93.73 \pm 3.87$& $39.04 \pm 3.87$& $690.95 \pm 17.63$& $286.00 \pm 27.73$\\
			R10 & -- & -- & -- &-- & -- &-- & -- & -- & --\\
			R11 & $ 57.80 \pm 19.29$& $77.15 \pm 23.81$& $100.00 \pm 2.10$& $100.00 \pm 8.30$& $8.59 \pm 0.55$& $8.21 \pm 0.84$& $47.81 \pm 2.43$ & $34.44 \pm 3.49$& $397.99 \pm 9.19$& $286.00 \pm 25.85$\\
			R12 & $ 49.35 \pm 5.31$&. $59.42\pm9.71$& $100.00 \pm 3.20$& $100.00 \pm 8.68$& $5.30 \pm 0.55$& $5.14 \pm 0.69$& $30.37 \pm 4.70$ & $24.60 \pm 4.32$ & $353.68 \pm 14.04$& $286.00 \pm 31.73$ \\
			R13 & $80.49 \pm 4.08$ & $93.88 \pm 13.98$& $100.00 \pm 1.20$& $100.00 \pm 9.16$& $18.07 \pm 0.67$& $17.64 \pm 2.29$& $37.29 \pm 4.66$ & $31.31 \pm 4.43$ & $341.02 \pm 9.18$& $286.00 \pm 28.90$\\
			R14 & $115.59 \pm 2.50$ & $151.47 \pm 17.87$& $100.00 \pm 1.29$& $100.00 \pm 6.85$& $17.54 \pm 0.49$& $16.81 \pm 1.50$& $52.78 \pm 4.80$ & $38.83 \pm 5.06$& $389.60 \pm 9.62$& $286.00 \pm 25.53$ \\
			R15 & $134.84 \pm 1.45$ & $205.98 \pm 12.85$& $100.00 \pm 0.59$& $100.00 \pm 4.98$& $73.52 \pm 0.93$& $68.78 \pm 4.40$& $55.61 \pm 3.39$& $34.38 \pm 3.15$& $464.31 \pm 6.50$ & $286.00 \pm 16.62$\\
\enddata		
\end{deluxetable}
\end{rotatetable}

	\movetabledown=2.2in	
	\begin{rotatetable}
         \begin{deluxetable}{cccccccccc}	
	\tablecaption{Emission line measurements (relative to H$\beta$ = 100) for the LBT spectra. Line fluxes (F$_{\lambda}$) are extinction corrected using E(B-V) to calculate I$_{\lambda}$.\label{tab:flux2}}	
\tablehead{ \colhead{Target} & \multicolumn{2}{c}{ \centering $[NII]$ } & \multicolumn{2}{c}{$[SII]$} & \multicolumn{2}{c}{$[SII]$} &\colhead{$E(B-V)$} & \multicolumn{2}{c}{F($H\beta$)} \\ 
\colhead{} & \multicolumn{2}{c}{(6583.41 \r{A})} & \multicolumn{2}{c}{(6716.47 \r{A})} & \multicolumn{2}{c}{(6730.85 \r{A})} & \colhead{} & \multicolumn{2}{c}{} \\
\colhead{} & \colhead{F$_{\lambda}$} & \colhead{I$_{\lambda}$}& \colhead{F$_{\lambda}$} & \colhead{I$_{\lambda}$}& \colhead{F$_{\lambda}$} & \colhead{I$_{\lambda}$}& \colhead{} & \colhead{F$_{\lambda}$} & \colhead{I$_{\lambda}$}}
\startdata
			R1 & $133.09 \pm 6.30$ &$93.71 \pm 10.58 $ &$51.77 \pm 2.20$ &$35.66 \pm 4.00$  &$36.12 \pm 1.79$ &$24.82 \pm 2.58$ &$0.324 \pm 0.029$ &$4.99 \pm 0.17 $& $15.07 \pm 1.21$ \\
			R2 & $166.45 \pm 7.90$ & $102.82 \pm 16.15$ &$48.07 \pm 1.99$ &$28.81 \pm 4.17$  &$34.50 \pm 1.77$ &$20.61 \pm 3.17$ &$0.446 \pm 0.033$ &$24.56 \pm 0.66$ &$111.73 \pm 14.49$ \\
			R3 & -- & -- & -- &-- & -- &-- & -- & --  \\
			R4 & $176.06 \pm 6.35$ & $105.65 \pm 9.31$ &$43.66 \pm 1.43$& $25.37 \pm 2.41$& $31.81 \pm 1.42$ & $18.42 \pm 1.75$& $0.472 \pm 0.021$ & $14.61 \pm0.010$  &$72.81 \pm 4.63$\\
			R5 &$167.42 \pm 9.38$&$103.22 \pm 16.80$& $53.17 \pm 2.64$& $31.80 \pm 5.07$& $39.26 \pm 2.28$ & $23.40 \pm 3.89$ & $0.448 \pm 0.038$& $13.69 \pm 0.49$ & $62.66 \pm 8.46$ \\
			R6 &$122.48 \pm 6.58$ & $99.41 \pm 15.57$ & $48.24 \pm 2.46$ & $38.65 \pm 6.47$ & $35.25 \pm 2.09$ & $28.20 \pm 4.75$ & $0.193 \pm 0.046$& $2.29 \pm 0.09$ &$ 4.41 \pm 0.58$\\
			R7 &  $376.16 \pm 6.56$& $139.90 \pm 6.99$ & $87.81 \pm1.54$ & $30.69 \pm 1.48$ & $86.61 \pm 1.53$ & $30.08 \pm 1.61$ & $0.915 \pm 0.012$ &  $32.37 \pm 0.30$ &  $726.68 \pm 29.03$ \\
			R8 & $226.87 \pm 9.19$ & $106.74 \pm 14.32$ & $62.44 \pm 2.17$ & $28.02 \pm 3.65$ & $ 44.84 \pm 1.92$ & $20.02 \pm 2.74$& $0.698 \pm 0.022$ &$3.58 \pm 0.08$& $38.41 \pm 4.70$  \\
			R9 &$287.21 \pm 7.84$ & $117.84 \pm 12.00$ & $73.17 \pm 1.94$  & $28.39 \pm 2.79$ & $55.09 \pm 1.54$ & $21.25 \pm 1.99$ & $0.824 \pm 0.021$ &$10.51 \pm 0.26$   & $173.18 \pm 14.73$\\
			R10 & -- & -- & -- &-- & -- &-- & -- & --\\
			R11 & $141.14 \pm 3.84$& $101.09 \pm 9.18$& $38.13 \pm 0.97$& $26.74 \pm 2.44$& $28.53 \pm 0.81$& $19.97 \pm 1.82$& $0.309 \pm 0.020$& $19.35 \pm 0.41$ & $55.29 \pm 4.59$ \\
			R12 &$90.05 \pm 5.68$& $72.66 \pm 8.56$&  $21.32 \pm 1.19$& $16.97 \pm 1.93$& $16.52 \pm 1.12$& $13.13 \pm 1.59$& $0.198 \pm 0.030$&  $29.15 \pm 0.93$ & $57.24 \pm 4.96$ \\
			R13 & $114.11 \pm 5.30$& $95.53 \pm 11.85$& $42.35 \pm 1.64$ & $35.06 \pm 3.87$ &  $29.67 \pm 1.58$& $24.54 \pm 2.90$&  $0.164 \pm 0.024$& $11.50 \pm 0.14$& $20.10 \pm 1.84$ \\
			R14 & $161.02 \pm 5.82$& $117.84 \pm 9.76$&$38.67 \pm 1.20$& $27.75 \pm 2.44$ & $28.24 \pm 1.16$& $20.22 \pm 1.72$& $0.289 \pm 0.020$ & $33.96 \pm 0.44$ & $90.67 \pm 6.21$    \\
			R15 & $171.92 \pm 3.91$& $105.38 \pm 6.28$ & $42.21 \pm 0.80$ & $25.09 \pm 1.37$  & $30.61 \pm 0.78$& $18.14 \pm 1.19$& $0.452 \pm 0.012$& $22.75 \pm 0.13$ & $106.07 \pm 5.28$  \\
\enddata
\tablecomments{F(H$\beta$) in units of $\times$ 10$^{-16}$ erg cm$^{-2}$ s$^{-1}$.}		
\end{deluxetable}
\end{rotatetable}

	\movetabledown=2.2in	
	\begin{rotatetable}
         \begin{deluxetable}{ccccccccccc}	
	\tablecaption{Emission line measurements (relative to H$\beta$ = 100) for MUSE spectra. Line fluxes (F$_{\lambda}$) are extinction corrected using E(B-V) to calculate I$_{\lambda}$. M83-2 is not included here because emission lines such as H$\beta$ and [OIII] 5007 were not detected with S/N $>3$, therefore we can not present emission line flux estimates relative to H$\beta$ and can not correct for dust-extinction.\label{tab:flux3}}	
\tablehead{ \colhead{Target} & \multicolumn{2}{c}{$H\beta$} & \multicolumn{2}{c}{ \centering $[OIII]$ } &\multicolumn{2}{c}{$[OIII]$} &\multicolumn{2}{c}{$[NII]$} & \multicolumn{2}{c}{$H\alpha$} \\ 
\colhead{} & \multicolumn{2}{c}{(4861.33 \r{A})} & \multicolumn{2}{c}{(4958.92 \r{A})} & \multicolumn{2}{c}{(5006.84 \r{A})} & \multicolumn{2}{c}{(6548.03 \r{A})} & \multicolumn{2}{c}{(6562.80 \r{A})} \\
\colhead{} & \colhead{F$_{\lambda}$} & \colhead{I$_{\lambda}$}& \colhead{F$_{\lambda}$} & \colhead{I$_{\lambda}$}& \colhead{F$_{\lambda}$} & \colhead{I$_{\lambda}$}& \colhead{F$_{\lambda}$} & \colhead{I$_{\lambda}$}& \colhead{F$_{\lambda}$} & \colhead{I$_{\lambda}$}}
\startdata
			M83-POS1 & $100.00 \pm0.43$ &$100.00 \pm 2.93$ & $2.82 \pm 0.29$ & $2.70\pm0.30$ & $7.86 \pm 0.46$ & $7.34 \pm 0.49$ & $78.30 \pm 3.28$ & $47.94 \pm 2.54$ & $468.80 \pm 5.81$&  $286.00 \pm 10.68$   \\
			M83-POS2 & $100.00 \pm 3.13$ & $100.00 \pm 8.80$ & -- & -- & $15.38 \pm 3.54$ & $14.60 \pm 2.78$ &    $63.62 \pm 3.75$   & $43.69 \pm 5.61$ & $417.61 \pm 14.15$ & $286.00 \pm32.49$   \\
			M83-1& $100.00 \pm 5.48$ & $100.00 \pm 9.75$ & -- &-- & -- &-- & $37.97 \pm 2.71$ & $22.80 \pm3.53$ & $478.10 \pm 26.41$ & $286.00 \pm 37.04$  \\
			M83-2 &  -- & -- & -- &-- & -- &-- & -- & -- \\
			M83-3 & $100.00 \pm 1.93$ & $100.00 \pm17.11$ & $11.67 \pm 1.54$ & $11.25 \pm 3.42$ & $24.65 \pm 2.24$ & $23.31 \pm 6.20$ & -- & -- & $429.41 \pm 22.35$ & $286.00 \pm 65.06$    \\
			M83-4& $100.00 \pm 1.51$ & $100.00 \pm 9.13$ & $8.81\pm1.13$ & $8.42\pm1.77$ & $19.90 \pm 1.69$ & $18.59 \pm 2.52$ &  $69.80 \pm 4.13$ & $42.80 \pm 5.41$ & $468.08 \pm 9.79$ & $286.00 \pm 28.28$  \\
			M83-5 & $100.00 \pm 1.80$ & $100.00 \pm 12.35$ & $7.89\pm1.40$ & $7.37\pm1.78$ & $15.59 \pm 1.97$ & $14.08 \pm 2.82$ & $87.48 \pm 6.69$ & $41.83 \pm 6.38$ & $601.30 \pm 15.53$& $286.00 \pm 38.29$     \\
			M83-6 & $100.00 \pm 10.03$ & $100.00 \pm 42.03$ &  -- & -- & -- & -- & $55.37 \pm 7.08$ & $26.64 \pm 15.25$ & $597.47 \pm 60.47$ & $286.00 \pm 151.10$ \\
			M83-10 &$100.00 \pm 1.82$ & $100.00 \pm 4.45$ & $5.69\pm1.23$ & $5.52\pm0.93$ & $12.87 \pm 1.83$ & $12.27 \pm 1.90$ & $37.52 \pm 1.95$ & $26.70 \pm 2.12$ & $402.91 \pm 8.05$ & $286.00 \pm 19.02$    \\
			M83-12 & $100.00 \pm 3.87$ & $100.00 \pm 9.62$& -- &-- & $13.16 \pm 3.94$& $12.31 \pm 4.08$ & $42.06 \pm 1.93$ & $25.97 \pm 3.12$ & $464.87 \pm 18.08$ & $286.00 \pm 31.98$ \\
			M83-14 & $100.00 \pm 1.44$ & $100.00 \pm 4.08$ & $4.26\pm0.95$ & $4.14\pm1.00$ & $10.68 \pm1.46$ & $10.21\pm 1.41$ &  $41.78 \pm 1.21$ & $30.29 \pm 1.76$ & $395.38 \pm 5.99$ & $286.00 \pm 17.25$  \\
			M83-16& $100.00 \pm 3.03$ & $100.00 \pm 9.39$ & $8.83\pm2.44$ & $8.50\pm3.40$ & $15.00 \pm 3.23$ & $14.16 \pm 2.40$ & $50.26 \pm 1.68$ & $33.23 \pm 3.40$ & $433.81 \pm13.20$ & $286.00 \pm 31.17$\\
\enddata		
\end{deluxetable}
\end{rotatetable}

	\movetabledown=2.2in	
	\begin{rotatetable}
         \begin{deluxetable}{cccccccccc}	
	\tablecaption{Emission line measurements (relative to H$\beta$ = 100) for MUSE spectra. Line fluxes (F$_{\lambda}$) are extinction corrected using E(B-V) to calculate I$_{\lambda}$. M83-2 is not included here because emission lines such as H$\beta$ and [OIII] 5007 were not detected with S/N $>3$, therefore we can not present emission line flux estimates relative to H$\beta$ and can not correct for dust-extinction.\label{tab:flux4}}	
\tablehead{ \colhead{Target} & \multicolumn{2}{c}{ \centering $[NII]$ } & \multicolumn{2}{c}{$[SII]$} & \multicolumn{2}{c}{$[SII]$} &\colhead{$E(B-V)$} & \multicolumn{2}{c}{F($H\beta$)} \\ 
\colhead{} & \multicolumn{2}{c}{(6583.41 \r{A})} & \multicolumn{2}{c}{(6716.47 \r{A})} & \multicolumn{2}{c}{(6730.85 \r{A})} & \colhead{} & \multicolumn{2}{c}{} \\
\colhead{} & \colhead{F$_{\lambda}$} & \colhead{I$_{\lambda}$}& \colhead{F$_{\lambda}$} & \colhead{I$_{\lambda}$}& \colhead{F$_{\lambda}$} & \colhead{I$_{\lambda}$}& \colhead{} & \colhead{F$_{\lambda}$} & \colhead{I$_{\lambda}$}}
\startdata
			M83-POS1 & $240.96 \pm 4.04$ & $146.28 \pm 5.86$ & $34.83 \pm0.76$ & $20.49 \pm 0.86$ & $38.22 \pm0.76$ & $22.41 \pm 0.93$  & $0.462 \pm 0.009$ & $1290.12\pm 5.51$ & $6201.11 \pm 181.94$\\
			M83-POS2 & $196.98 \pm 7.21$ & $134.39 \pm 15.17$ & $40.86 \pm 1.79$ & $27.22 \pm 3.05$ & $42.55 \pm 1.84$ & $28.27 \pm 3.04$ & $0.354 \pm 0.030$ & $616.45 \pm 19.31$ & $2052.09 \pm 180.54$\\
			M83-1& $118.61 \pm 6.76$ & $70.59 \pm 10.37$ & $48.12 \pm 2.94$ & $27.72 \pm 4.21$ & $38.96 \pm 2.51$ & $22.37 \pm 3.15$ & $0.480 \pm 0.042$ & $11.5 \pm 0.63$ & $58.85 \pm 5.74$   \\
			M83-2 & -- & -- & -- &-- & -- &-- & -- & -- & -- \\
			M83-3 & $232.51 \pm 21.16$ & $154.23 \pm 31.38$ & $51.78 \pm 1.99$ & $33.47 \pm 6.74$ &  $56.59 \pm2.05$ & $36.48 \pm7.01$ & $0.38 \pm 0.048$ &  $940.14 \pm 18.18$ & $3419.31 \pm 585.15$ \\
			M83-4& $215.40 \pm 5.69 $& $130.96 \pm 13.55$ & $51.98 \pm 1.67$ & $30.63 \pm 3.38$ & $53.78 \pm 1.68$ & $31.59 \pm 3.28$ & $0.460 \pm 0.022$ &  $854.20 \pm 12.86$ & $4085.82 \pm 373.02$  \\
			M83-5 & $270.34 \pm 9.01$  & $127.63 \pm 17.29$ & $57.17 \pm 1.87$ & $25.75 \pm 3.41 $ &$53.89 \pm 1.84$ & $24.15 \pm 3.44$ &  $0.694 \pm 0.029$ & $267.25 \pm 4.82$ & $2832.69 \pm 349.87$   \\
			M83-6 & $169.90 \pm 17.69$  & $80.73 \pm36.57$ & $49.27 \pm 5.29$ & $22.34 \pm 10.12$ & $40.53 \pm 4.50$ & $18.29 \pm 8.79$ & $0.688 \pm 0.112$ & $12.95 \pm 1.3$ & $134.49 \pm 56.53$  \\
			M83-10 & $111.70 \pm 3.10$ & $79.02 \pm 5.30$ & $35.87 \pm 0.77$ & $24.83 \pm 1.75$ & $26.13 \pm 0.62$ & $18.04 \pm 1.11$ &$0.320 \pm 0.021$ & $46.57 \pm 0.85$ & $138.34 \pm 6.16$   \\
			M83-12 & $126.50 \pm 5.02$ & $77.45 \pm 9.36$ & $48.51 \pm 2.00$ & $28.80 \pm 3.18$  & $36.44 \pm 1.56$ & $21.56 \pm 2.36$  & $0.454 \pm 0.029$ & $22.86 \pm 0.88$ & $106.98 \pm 10.29$ \\
			M83-14 &  $126.77 \pm 2.16$& $91.40 \pm 5.00$ & $46.77 \pm0.89$ & $33.04 \pm 1.78$ &  $33.51 \pm 0.77$ & $23.62 \pm 1.32$ & $0.303 \pm 0.014$ & $59.50 \pm 0.86$ & $166.48 \pm 6.78$  \\
			M83-16& $157.12 \pm4.83$ & $103.16 \pm 10.85$ &  $65.62 \pm2.09$ & $41.96 \pm 4.47$ &  $47.44 \pm1.57$ & $30.25 \pm 3.11$ &  $0.389 \pm 0.023$ & $26.69 \pm 0.81$ & $100.27 \pm 9.41$  \\
\enddata
\tablecomments{F(H$\beta$) in units of $\times$ 10$^{-16}$ erg cm$^{-2}$ s$^{-1}$.}		
\end{deluxetable}
\end{rotatetable}




\bibliographystyle{aasjournal}
\bibliography{M83} 


\end{document}